\documentclass{aa}

\usepackage[varg]{txfonts}

\begin{document}
    \title{The MUSE-Faint survey: I.~Spectroscopic evidence for a star cluster in Eridanus~2 and constraints on MACHOs as a constituent of dark matter\thanks{Based on observations made with ESO Telescopes at the La Silla Paranal Observatory under programme ID 0100.D-0807.}}
    \titlerunning{MUSE-Faint: I.~Spectroscopy on the Eri~2 cluster and constraints on MACHO DM}
    \author{%
        Sebastiaan~L. Zoutendijk\inst{1} \and
        Jarle Brinchmann\inst{2,1} \and
        Leindert~A. Boogaard\inst{1} \and
        Madusha~L.~P. Gunawardhana\inst{1} \and
        Tim-Oliver Husser\inst{3} \and
        Sebastian Kamann\inst{4} \and
        Andr\'es Felipe Ramos Padilla\inst{1}\fnmsep\thanks{Current address: Kapteyn Astronomical Institute, University of Groningen, P.O.~Box~800, 9700~AV~Groningen, The Netherlands.} \and
        Martin M. Roth\inst{5} \and
        Roland Bacon\inst{6} \and
        Mark den Brok\inst{5} \and
        Stefan Dreizler\inst{3} \and
        Davor Krajnovi\'c\inst{5}%
    }
    \authorrunning{S.~L.~Zoutendijk et al.}
    \institute{%
        Leiden Observatory, Leiden University, P.O.~Box~9513, 2300~RA~Leiden, The Netherlands\\\email{zoutendijk@strw.leidenuniv.nl} \and
        Instituto de Astrof\'{\i}sica e Ci\^encias do Espa\c{c}o, Universidade do Porto, CAUP, Rua das Estrelas, PT4150-762~Porto, Portugal \and
        Institute for Astrophysics, Georg-August-University, Friedrich-Hund-Platz~1, 37077~G\"ottingen, Germany \and
        Astrophysics Research Institute, Liverpool John Moores University, 146~Brownlow Hill, Liverpool~L3~5RF, United Kingdom \and
        Leibniz-Institut f\"ur Astrophysik Potsdam (AIP), An der Sternwarte 16, D-14482 Potsdam, Germany \and
        Univ. Lyon, Univ. Lyon1, ENS de Lyon, CNRS, Centre de Recherche Astrophysique de Lyon UMR5574, 69230, Saint-Genis-Laval, France%
    }
    \date{Received date / Accepted date}
    \abstract{%
    }{%
        It has been shown that the ultra-faint dwarf galaxy Eridanus~2 may host a stellar cluster in its centre.
        This cluster, if shown to exist, can be used to set constraints on the mass and abundance of massive astrophysical compact halo objects~(MACHOs) as a form of dark matter.
        Previous research has shown promising expectations in the mass range of $10$--$100\,M_\mathrm{\sun}$, but lacked spectroscopic measurements of the cluster.
        We aim to provide spectroscopic evidence regarding the nature of the putative star cluster in Eridanus~2 and to place constraints on MACHOs as a constituent of dark matter.%
    }{%
        We present spectroscopic observations of the central square arcminute of Eridanus~2 from MUSE-Faint, a survey of ultra-faint dwarf galaxies with the Multi Unit Spectroscopic Explorer on the Very Large Telescope.
        We derive line-of-sight velocities for possible member stars of the putative cluster and for stars in the centre of Eridanus~2.
        We discuss the existence of the cluster and determine new constraints for MACHOs using the Fokker--Planck diffusion approximation.%
    }{%
        Out of 182~extracted spectra, we identify 26 member stars of Eridanus~2, seven of which are possible cluster members.
        We find intrinsic mean line-of-sight velocities of $79.7^{+3.1}_{-3.8}\,\mathrm{km}\,\mathrm{s}^{-1}$ and $76.0^{+3.2}_{-3.7}\,\mathrm{km}\,\mathrm{s}^{-1}$ for the cluster and the bulk of Eridanus~2, respectively, and intrinsic velocity dispersions of ${<}7.6\,\mathrm{km}\,\mathrm{s}^{-1}$ (68-$\%$ upper limit) and $10.3^{+3.9}_{-3.2}\,\mathrm{km}\,\mathrm{s}^{-1}$, respectively.
        This indicates the cluster most likely exists as a distinct dynamical population hosted by Eridanus~2, without surplus of dark matter over the background distribution.
        Among the member stars in the bulk of Eridanus~2, we find possible carbon stars, alluding to the existence of an intermediate-age population.
        We derive constraints on the fraction of dark matter that can consist of MACHOs with a given mass between $1$--$10^5\,M_\sun$.
        For dark matter consisting purely of MACHOs, the mass of the MACHOs must be less than ${\sim}7.6\,M_\sun$ and ${\sim}44\,M_\sun$ at a $68$- and $95$-$\%$ confidence level, respectively.%
    }{%
    }
    \keywords{dark matter -- galaxies: individual: \object{Eridanus~2} -- galaxies: star clusters: individual: \object{Eridanus~2 cluster} -- techniques: imaging spectroscopy}
    \maketitle

\section{Introduction}
\label{sec:introduction}
    Ultra-faint dwarf galaxies~(UFDs), defined as dwarf galaxies with an absolute V-band magnitude of $M_\mathrm{V} > -7.7$~\citep{Simon-2019-ARA&A-57-375}, are the faintest galaxies that we can currently observe.
    First discovered in~2005~\citep{Willman-2005-AJ-129-2692, Willman-2011-AJ-142-128}, astronomers have found about 30~confirmed and 40~candidate UFDs\footnote{The exact numbers of confirmed and candidate UFDs depend on the exact definition of a UFD and on the exact criteria for confirmation.} since then, mostly through analysis of wide-field image surveys such as the Sloan Digital Sky Survey~\citep[SDSS;][]{York-2000-AJ-120-1579} and the Dark Energy Survey~\citep[DES;][]{Abbott-arXiv-0510}.
    While more luminous dwarf galaxies can be easily distinguished from globular clusters~(GCs) through their location in the magnitude--half-light radius plane, photometric classification becomes ambiguous below $M_\mathrm{V} \sim -6$ \citep[see e.g.][]{Koposov-2015-ApJ-805-130}.
    Spectroscopy can break this degeneracy because, contrary to GCs, the stars in UFDs are embedded in a dark-matter halo~\citep[see e.g.][]{Willman-2012-AJ-144-76}.
    This will be visible in measurements of velocity dispersion and indirectly through the metallicity distribution.

    The high dark-matter content of UFDs is in fact what makes them interesting to cosmology.
    The dynamical mass--to--light ratios of UFDs are the highest measured in any galaxy \citep[see e.g.][]{McConnachie-2012-AJ-144-4}.
    Since this implies relatively pure dark-matter haloes with minimal baryonic content, UFDs are an ideal target for dark-matter studies.
    One area where UFDs can advance our knowledge of dark matter is the so-called core--cusp problem \citep[see e.g.][]{Bullock-2017-ARA&A-55-343}.
    While dark matter--only simulations predict the densities of galaxies to follow a Navarro--Frenk--White profile~\citep{Navarro-1996-ApJ-462-563, Navarro-1997-ApJ-490-493}, which has a cusp at the centre, observations of larger dwarf galaxies show the presence of a core with constant density~\citep{Brooks-2014-ApJ-786-87, DiCintio-2014-MNRAS-437-415}.
    If one believes these cores are the result of supernova feedback between baryons and dark matter, it is predicted that UFDs will not show a core, as they have not formed enough stars to have significant supernova feedback~\citep{Pennarrubia-2012-ApJ-759-L42}.
    The latest high-resolution hydrodynamical simulations, which can now resolve the internal structure of UFDs, show this as well~\citep{Onnorbe-2015-MNRAS-454-2092, Wheeler-2019-MNRAS-490-4447}.
    Observations of cuspy UFD density profiles will therefore support this theory of baryonic feedback and indicate that there is no need to revise our theories on dark matter in this respect.
    On the other hand, cored UFDs would indicate either a lack of understanding of baryonic feedback processes, or a different nature of dark matter than assumed in the simulations.
    For instance, simulations of somewhat larger dwarf galaxies indicate that self-interacting dark matter with or without baryonic feedback can cause a cored density profile even when cold dark matter with or without baryonic feedback cannot~\citep{Fitts-2019-MNRAS-490-962}.

    The UFD Eridanus~2~(Eri~2; also known as Eridanus~II and DES J0344.3--4331) has another characteristic that makes it an interesting object to study.
    It likely contains a stellar cluster, most likely near the centre~\citep{Crnojevic-2016-ApJL-824-L14, Contenta-2018-MNRAS-476-3124}, and possibly globular~\citep{Koposov-2015-ApJ-805-130}.
    We summarize the properties relevant for this paper that have been determined in previous studies for both the bulk of Eri~2 and its putative cluster in Table~\ref{tab:prop}.
    \begin{table}
        \caption{%
            Relevant properties of Eridanus 2 and its putative cluster, known from previous studies.
            We list the positions on the sky, absolute $\mathrm{V}$-band magnitudes, heliocentric distance, projected half-light radii, intrinsic line-of-sight systemic velocity and velocity dispersion, half-light mass, mass-to-light ratio, mean metallicity, and metallicity dispersion.
        }
        \label{tab:prop}
        \centering
        \begin{tabular}{lcc}
            \hline\hline
            Parameter & Eridanus 2 & Putative cluster \\
            \hline
            Right ascension & $03\mathrm{h}44\mathrm{m}20\fs1 \pm 10\fs5$ & $03\mathrm{h}44\mathrm{m}22\fs2 \pm 1\mathrm{s}$ \\
            Declination & $-43\degr32\arcmin01\farcs7 \pm 5\farcs3$ & $-43\degr31\arcmin59\farcs2 \pm 2\arcsec$ \\
            $M_\mathrm{V}$ ($\mathrm{mag}$) & $-7.1 \pm 0.3$ & $-3.5 \pm 0.6$ \\
            $D$ ($\mathrm{kpc}$) & $366 \pm 17$ & --- \\
            $R_\mathrm{h}$ ($\mathrm{pc}$) & $277 \pm 14$ & $13 \pm 1$ \\
            \hline
            $\mu_\mathrm{int}$ ($\mathrm{km}\,\mathrm{s}^{-1}$) & $75.6 \pm 1.3 \pm 2.0$\tablefootmark{a} & --- \\
            $\sigma_\mathrm{int}$ ($\mathrm{km}\,\mathrm{s}^{-1}$) & $6.9^{+1.2}_{-0.9}$ & --- \\
            $M_\mathrm{h}$ ($M_\sun$) & $1.2^{+0.4}_{-0.3} \times 10^7$ & --- \\
            $M_\mathrm{h}/L_\mathrm{V}$ ($M_\sun/L_\sun$) & $420^{+210}_{-140}$ & --- \\
            $\mu_{[\mathrm{Fe/H}]}$ ($\mathrm{dex}$) & $-2.38 \pm 0.13$ & --- \\
            $\sigma_{[\mathrm{Fe/H}]}$ ($\mathrm{dex}$) & $0.47^{+0.12}_{-0.09}$ & --- \\
            \hline
        \end{tabular}
        \tablefoot{%
            The first five parameters are taken from the photometric study by \citet{Crnojevic-2016-ApJL-824-L14}.
            The last six parameters were determined by \citet{Li-2017-ApJ-838-8} based on spectroscopy at large radii.
            \tablefoottext{a}{The two uncertainties given are random respectively systematic.}
        }
    \end{table}
    If the observed stellar overdensity is indeed a cluster, it can be used as a cosmological probe, in (at least) three ways.

    This putative cluster could offer good constraints on the abundance of massive astrophysical compact halo objects~(MACHOs; \citealt{Griest-1991-ApJ-366-412}) as a form of dark matter~\citep{Brandt-2016-ApJL-824-L31, Li-2017-ApJ-838-8, Zhu-2018-MNRAS-476-2}.
    Such objects would transfer kinetic energy to the stars in the cluster through dynamical interactions.
    This would cause the stars in the cluster to move to wider orbits, eventually dissolving the cluster.
    The survival of the cluster, therefore, constrains the strength of the interaction between MACHOs and stars, which is a function of MACHO mass and abundance.
    To set these constraints the existence of the putative cluster has to be established and radial velocities of stars in the cluster and in the bulk of Eri~2 are required, but so far spectroscopic observations of the putative star cluster have been lacking.
    Of particular cosmological interest is the work by \citet{Brandt-2016-ApJL-824-L31}, predicting that constraints could be placed on the abundance of MACHOs with masses of $10$--$100\,M_\sun$ using a diffusion approximation.
    It has been proposed~\citep{Clesse-2015-PhRvD-92-023524, Bird-2016-PhRvL-116-201301} that mergers of black holes with masses ${\sim}30\,M_\sun$ --~such as observed~\citep[][though see also \citealt{Broadhurst-arXiv-1802}]{Abbott-2016-PhRvL-116-061102} with the Laser Interferometer Gravitational-Wave Observatory~(LIGO)~--, which are higher masses than expected for stellar black holes, could be from primordial black holes \citep{Hawking-1971-MNRAS-152-75, Carr-1974-MNRAS-168-399}, possibly making up an appreciable fraction of dark matter.
    Some constraints have already been placed in this mass range~(see \citealt{Carr-2017-PhRvD-96-023514} and references therein; their Fig.~1 includes the constraints expected for Eri~2 as calculated by~\citealt{Brandt-2016-ApJL-824-L31}), but every constraint has its weaknesses due to the assumptions taken in its derivation.
    It therefore remains interesting to place another, independent constraint based on the putative cluster in Eri~2.
    Even though large parts of the parameter space for MACHOs are already ruled out and other dark-matter candidates are usually preferred by cosmologists, this should not stop us from trying to place further constraints when we can, as advocated by \citet{Bertone-2018-Natur-562-51}.

    Furthermore, constraints can be placed on the mass and abundance of ultra-light dark-matter particles using a similar analysis~\citep{Marsh-2019-PhRvL-123-051103}.
    In this theory of dark matter, density perturbations occur in the centre of a dark-matter halo.
    These perturbations can be treated in a statistically similar manner to heating due to MACHOs.

    Finally, the fact that the putative cluster has survived in the centre of a dark-matter halo can put constraints on the dark-matter density profile in the centre of this halo~\citep{Li-2017-ApJ-838-8, Contenta-2018-MNRAS-476-3124}.
    If the cluster is near the centre of the dark-matter halo, and if this halo's density profile had too large a cusp, the cluster would have been destroyed.

    Eri~2 was independently discovered by \citet{Koposov-2015-ApJ-805-130} in public DES data and by \citet{Bechtol-2015-ApJ-807-50} in DES internal data release Y1A1.
    Based on deep imaging with the Megacam on the Magellan Clay Telescope, an absolute magnitude $M_\mathrm{V} = -7.1\pm0.3$, a projected half-light radius $R_{\mathrm{h},\mathrm{gal}} = 277\pm14\,\mathrm{pc}$, and a distance of ${\sim}366\pm17\,\mathrm{kpc}$ were determined~\citep{Crnojevic-2016-ApJL-824-L14}.
    Using the IMACS spectrograph on the Magellan Baade Telescope, \citet{Li-2017-ApJ-838-8} measured a velocity dispersion of $\sigma = 6.9^{+1.2}_{-0.9}\,\mathrm{km}\,\mathrm{s}^{-1}$ and inferred a mass-to-light ratio of $420^{+210}_{-140}\,M_\sun\,L_\sun^{-1}$, confirming Eri~2 is an UFD.
    The earliest observations indicated the possible presence of bright blue stars and a stellar cluster~\citep{Koposov-2015-ApJ-805-130}, and an age of ${\sim}12\,\mathrm{Gyr}$ for the main population~\citep{Bechtol-2015-ApJ-807-50}.
    The Megacam imaging showed the blue stars were likely an intermediate-age population and confirmed the presence of a cluster-like overdensity.
    The spectroscopy did not prove the existence of an intermediate-age population, but could not rule it out either.
    No spectroscopic observations were made of the putative cluster, so it remains unconfirmed whether this is a cluster or some other overdensity and whether it belongs to Eri~2 or is only seen near it in projection.

    Spectroscopic observations of the stellar overdensity are needed to establish its nature and determine its dynamical properties in order to constrain MACHOs as a significant form of dark matter in the mass range $1$--$10^5\,M_\sun$.
    In this paper, we present spectroscopic observations of the central square arcminute of Eri~2 from MUSE-Faint, a survey of UFDs with the Multi Unit Spectroscopic Explorer~(MUSE) on the Very Large Telescope~(VLT).
    Using the line-of-sight velocities of carefully selected member stars, we characterize the velocity distributions of stars in the putative star cluster and the bulk of Eri~2.
    We show that the star cluster likely exists, and use its survival to constrain its dynamical heating due to MACHO dark matter using a diffusion approximation, thereby placing constraints on the fraction of MACHOs in dark matter.
    We also discuss within what limits the diffusion approximation is expected to be valid.

    In Sect.~\ref{sec:reduction} we summarize the observations (Sect.~\ref{ssec:obs}) and describe the data reduction process, first the steps to obtain a data cube (Sect.~\ref{ssec:cube}) and then the extraction of the spectra (Sect.~\ref{ssec:spectra}).
    We then continue with the astrophysical results in Sect.~\ref{sec:astrophys}, showing the determination of line-of-sight velocities from the spectra (Sect.~\ref{ssec:velocities}), fitting distributions to these velocities (Sect.~\ref{ssec:distributions}), and inferring a number of properties of Eri~2 and its putative star cluster (Sect.~\ref{ssec:properties}).
    This is followed by the implications for MACHOs in Sect.~\ref{sec:impl}, where we first present a mathematical model for MACHOs that can be applied to our data (Sect.~\ref{ssec:model}) and then constrain MACHO mass and fraction of dark matter using this model and the velocity distributions (Sect.~\ref{ssec:constraints}).
    We close with a discussion (Sect.~\ref{sec:discussion}) and conclusions (Sect.~\ref{sec:conclusions}).
    In Appendix~\ref{app:fMlim} we show the derivation of the model presented in Sect.~\ref{ssec:model}.

\section{Observations and data reduction}
\label{sec:reduction}

\subsection{Observations}
\label{ssec:obs}
    We observed the central region of Eri~2 with MUSE~\citep{Bacon-2010-SPIE-7735-773508} on Unit Telescope~4 of the VLT.
    MUSE is an integral-field spectrograph with a spectral sampling of $1.25\,\mathrm{\AA}$, a spectral resolution with a full width at half-maximum~(FWHM) between $2.4\,\mathrm{\AA}$ and $3.0\,\mathrm{\AA}$~\citep{Bacon-2017-A&A-608-A1}, and a nominal wavelength coverage of $4800$--$9300\,\mathrm{\AA}$.
    We used MUSE in the wide-field mode with adaptive optics~(AO) and nominal read-out mode, providing a $1\,\mathrm{arcmin} \times 1\,\mathrm{arcmin}$ field of view and a $0.2\,\mathrm{arcsec} \times 0.2\,\mathrm{arcsec}$ spatial sampling.
    The relatively large field of view for an integral-field spectrograph makes MUSE an ideal instrument to study UFDs.
    One can capture medium-resolution spectra of tens or hundreds of stars within the half-light radius of an UFD with only a few pointings and without the need for pre-selection of targets.

    The exposures total $4.5\,\mathrm{h}$ and were taken during Guaranteed Time Observing~(GTO) runs between 16 and 25~October 2017 ($3\,\mathrm{h}$) and between 14 and 19~March 2018 ($1.5\,\mathrm{h}$) under programme ID~0100.D-0807.
    The data were taken in 900-s exposures in groups of three or four, where each of the exposures is rotated $90\,\mathrm{deg}$ relative to the previous exposure and dithered.
    The VLT auto-guider indicated a natural seeing of around $0.7\,\mathrm{arcsec}$ during the first run, while the slow--guide star system showed an AO-corrected seeing of ${\sim}0.5\,\mathrm{arcsec}$.
    The second run had a natural seeing of ${\sim}0.8\,\mathrm{arcsec}$, which was not improved by AO.
    After reduction we measure a FWHM of $0.528\,\mathrm{arcsec}$ at a wavelength of $7000\, \mathrm{\AA}$.

\subsection{Data cube}
\label{ssec:cube}
    We processed the raw MUSE data, in the form of one detector image per integral-field unit per exposure, into a data cube, using calibration files provided by the MUSE GTO consortium.
    Unless otherwise noted, the data were processed with EsoRex using the MUSE Data Reduction Software~\citep[DRS; version~2.4;][]{Weilbacher-2012-SPIE-8451-84510B}.

    First, we performed basic science processing on each exposure, applying the relevant master bias, master flat, trace table, wavelength calibration table, geometry table, and twilight cube.
    We also applied an illumination exposure for each science exposure, to account for temperature differences between the science and calibration exposures.
    Of the two illumination exposures adjacent to the science exposure, we selected the one closest in ambient temperature.
    Furthermore, we applied the bad-pixel table created by \citet{Bacon-2017-A&A-608-A1}.
    After these steps we were left with one object pixel table for each of the 24 integral-field units and for each exposure.

    We then continued with science post-processing of the object pixel tables using the standard procedure for AO data.
    We enabled autocalibration using the deep-field method, an updated version of the method described by \citet{Bacon-2017-A&A-608-A1} and included in the DRS.
    The result was a reduced pixel table, a data cube, and a white-light image, for each exposure.
    We determined spatial offsets from the white-light images and combined the exposures into a single cube in two steps to reduce memory consumption.

\begin{figure*}
        \includegraphics[width=0.5\linewidth]{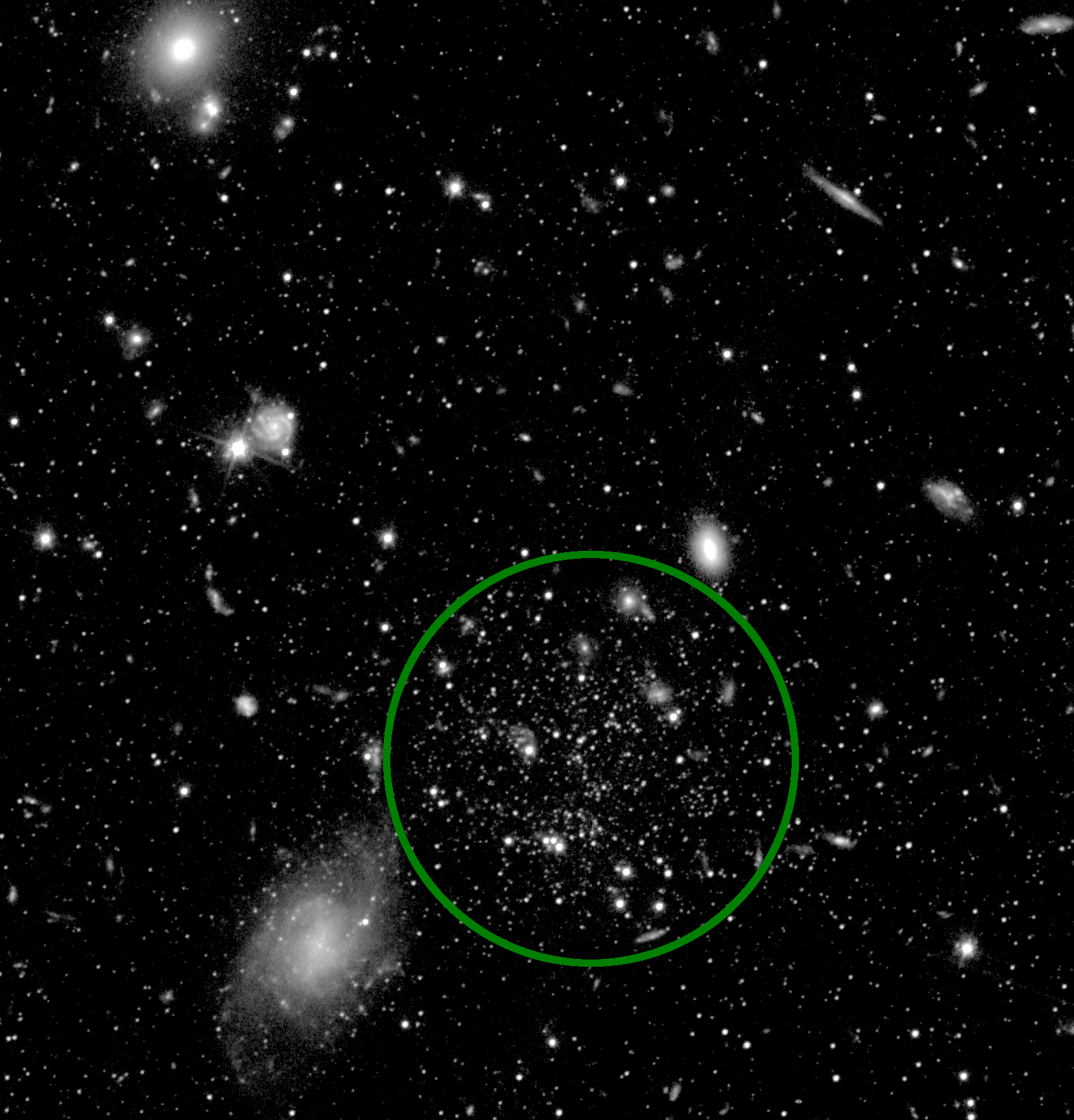}%
        \includegraphics[width=0.5\linewidth]{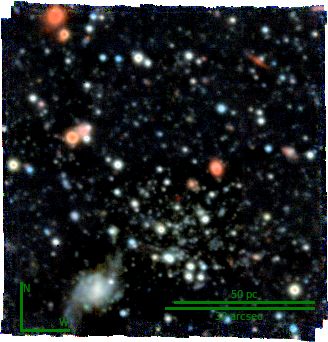}%
        \caption{%
            Images of the centre of Eridanus~2, shown at the same scale.
            \emph{Left:}
            Image of public \emph{Hubble Space Telescope} data, cut to the same size as our MUSE observations.
            The sources inside the green circle are possible members of the putative star cluster.
            \emph{Right:}
            Composite colour image of the same area seen with MUSE, using Cousins~R, Johnson~V, and Johnson~B filters for red, green, and blue, respectively.
            The angular and physical scales are indicated, as are north and west.
        }
        \label{fig:im}
    \end{figure*}

    Since the deep-field autocalibration relies on the sky flux, a good sky mask is necessary to achieve the best result.
    The observations discussed here are over a fairly filled field and the MUSE white-light image is insufficient to make a precise object-free sky mask.
    To overcome this problem, we first created a source catalogue from $27\,360\,\mathrm{s}$ of public \emph{Hubble Space Telescope}~(\emph{HST}) data of Eri~2\footnote{\emph{Hubble Space Telescope} Proposal~14234, principal investigator J.\,D.~Simon.}.
    An image of these data is shown in the left panel of Fig.~\ref{fig:im}.
    We first created a list of extended sources from the \emph{HST} image using SExtractor~\citep{Bertin-1996-A&AS-117-393} and masked them out.
    Then we ran DAOPHOT~II\footnote{As we used images normalized by exposure time, it was necessary to make small modifications to the source code.}~\citep{Stetson-1987-PASP-99-191} on the masked \emph{HST} image, to create a catalogue of point sources, extracting down to $5\sigma$ using an empirical point-spread function model.
    We then calibrated the \emph{HST} image's world coordinate system~(WCS) information to that of the white light--filtered combined MUSE image by visually matching sources in the images.
    Taking the extended sources from the SExtractor catalogue and the point sources from DAOPHOT~II, we created a single source catalogue and transformed it to the MUSE WCS.
    The catalogue's photometric zero point was calibrated according to the photometric calibration information in the \emph{HST} image header.
    The mask of extended objects underwent the same WCS transformation as the catalogue.
    We added single pixels at the positions of the detected point sources brighter than $27\,\mathrm{mag}$.
    This value was chosen because it appeared to be the sweet spot between removing as much excess flux as possible and minimizing the removal of pixels that do not show a flux excess above the sky level in the MUSE white-light image.
    The mask was then convolved with a 0.8-arcsec top-hat kernel to achieve a resolution similar to MUSE.

    The science post-processing and exposure combination were then repeated.
    In addition to the standard files, we also supplied to the science post-processing the offset list, the new mask as sky mask, and the previously combined data cube as output WCS, to ensure the mask is properly aligned with each exposure.
    We also set \verb|skymodel_fraction| to $0.95$ because the supplied mask should cover (almost) all of the sources.
    The resulting pixel tables were combined using the same settings as in the previous exposure combination.
    To remove residual sky signatures, we ran the Zurich Atmosphere Purge~\citep[ZAP; version~2.0;][]{Soto-2016-MNRAS-458-3210} on the new combined data cube, using the \emph{HST}-derived sky mask.
    A composite colour image of the cube, using Johnson B and V and Cousins R filters, is shown in the right panel of Fig.~\ref{fig:im}.

\subsection{Spectra}
\label{ssec:spectra}
\begin{figure*}
        \includegraphics[width=\linewidth]{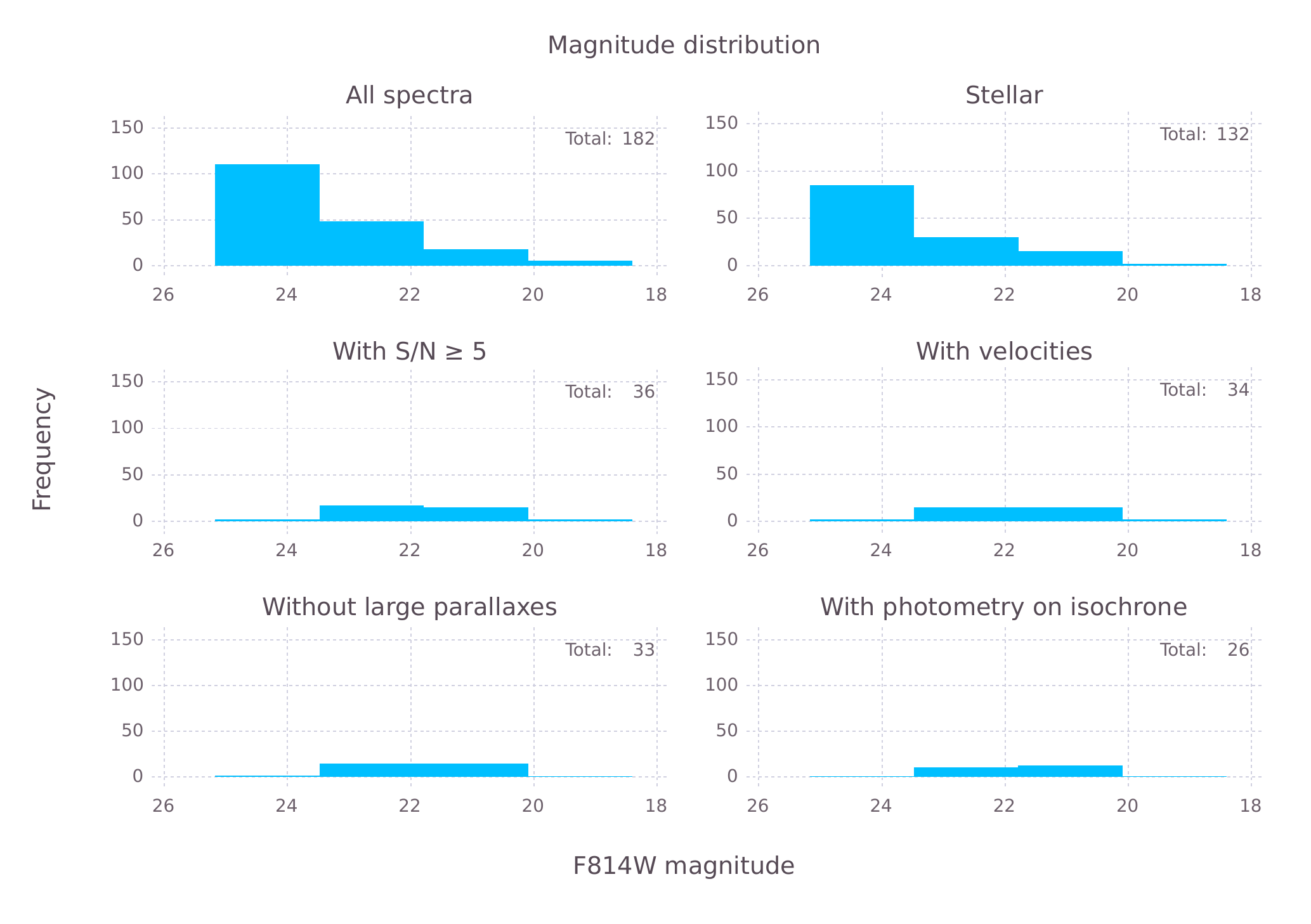}%
        \caption{%
            Magnitude distribution of sources after extraction and after application of selection criteria.
            The criteria applied cumulatively are plausible stellarity, a spectral signal-to-noise ratio of at least $5$, the presence of a radial velocity measurement, the lack of a large parallax, and photometry consistent with a broadened isochrone.
        }
        \label{fig:mag}
    \end{figure*}

    To make further progress we needed to extract the spectra of point sources in our data cube which may at times be blended with each other.
    To do this we used PampelMuse, which is described in detail by \citet{Kamann-2013-A&A-549-A71}, and the same source catalogue as used for masking.
    Briefly, it determines an initial point-spread function~(PSF) by fitting a Moffat function to a spectrally binned data cube (in our case, a spectral sampling of $62.5\, \mathrm{\AA}$), at the locations of PSF stars.
    A polynomial fit as a function of wavelength to the PSF star positions (to account for optical aberrations and inaccuracies in the differential atmospheric refraction correction) and PSF parameters $\beta$ and the FWHM (to account for the variation of the seeing) then provides an initial guess for a second fit to the original, non-binned data cube.
    We empirically chose third-order polynomials for the positions and linear fits for the PSF parameters on five PSF stars simultaneously.
    The values of the fitted parameters at a wavelength of $7000\, \mathrm{\AA}$ were $\beta = 2.61$ and a FWHM of $2.64$~pixels or $0.528\,\mathrm{arcsec}$.
    Taking into account the recovered PSF, we extracted the spectra of all sources where, before extraction, PampelMuse estimated a signal-to-noise ratio~(S/N) per spectral element of at least~$1$.
    As a result, we had spectra for 182~sources, excluding three spectra with a negative median value.
    The magnitude distribution of all extracted spectra is shown in Fig.~\ref{fig:mag}.
    The width of the bins of this and all other histograms in this paper are calculated using Knuth's rule~\citep{Knuth-arXiv-0605}, which maximizes the posterior probability that the histogram describes the distribution function underlying the data.

\section{Astrophysical results}
\label{sec:astrophys}
    Having extracted the spectra from the reduced data cube, we now continue with their analysis.
    First we determine line-of-sight velocities for each spectrum and determine which sources are member stars of Eridanus~2 (Sect.~\ref{ssec:velocities}).
    We then determine the intrinsic velocity distribution of Eridanus~2 and its putative cluster (Sect.~\ref{ssec:distributions}).
    Next we discuss the existence of the cluster, determine masses for Eridanus~2 and its cluster, and show the metallicity distribution of our sample (Sect.~\ref{ssec:properties}).

\subsection{Line-of-sight velocities}
\label{ssec:velocities}
\begin{figure*}
        \includegraphics[width=\linewidth]{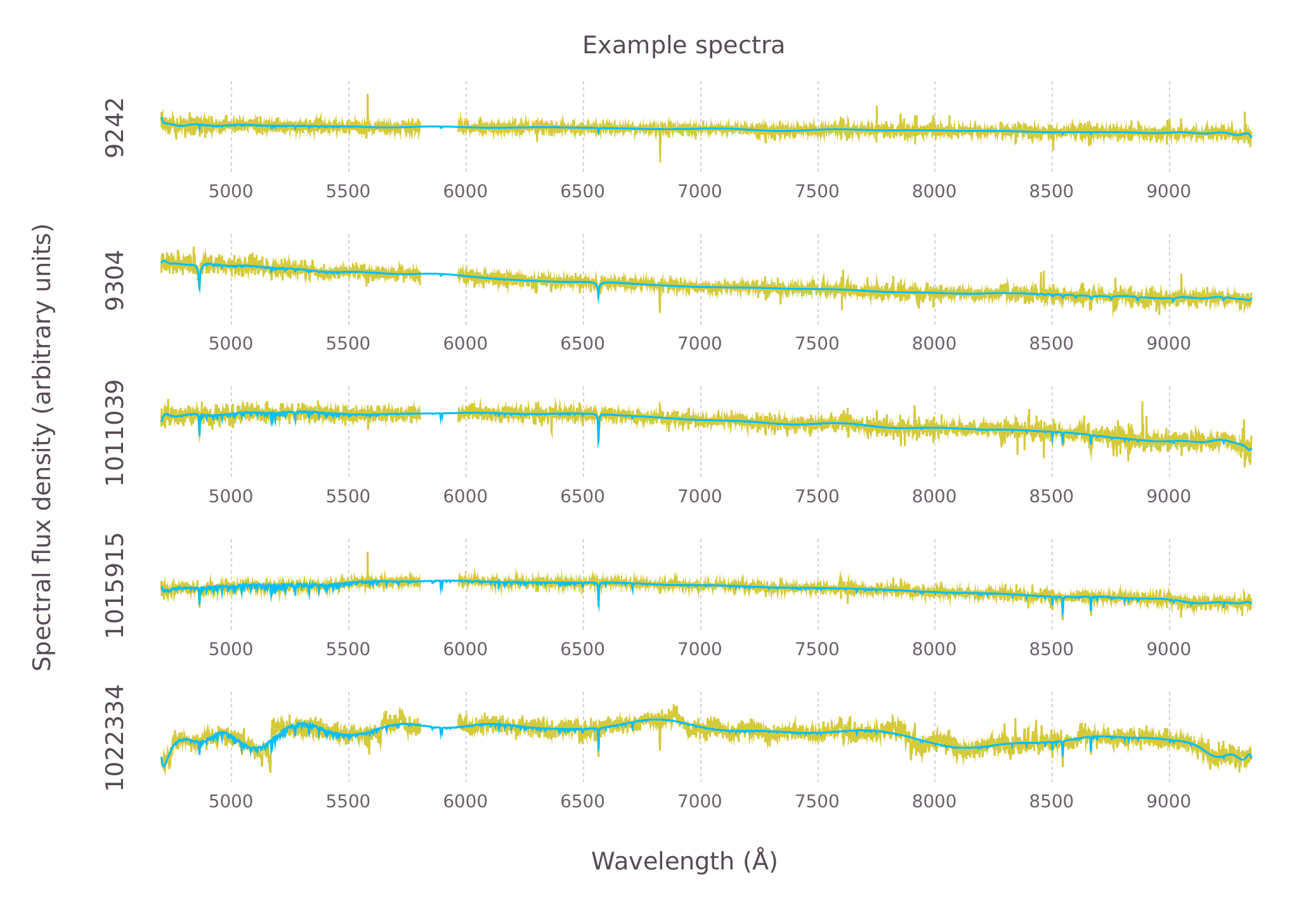}%
        \caption{%
            Illustration of typical spectra at different signal-to-noise levels, with source IDs shown on the left.
            These spectra have empirical signal-to-noise ratios of $6$, $8$, $17$, $28$, and $15$, from top to bottom.
            The extracted spectra are shown in yellow, with the best-fitting spectra found by spexxy overplotted in blue.
            The gap in the observed spectrum just blueward of $6000\,\mathrm{\AA}$ is masked because it contains the Na~D emission of the adaptive optics laser.
            The last source is a carbon star; as the PHOENIX library used with spexxy does not include these type of stars, the best-fitting spectrum matches the absorption lines but can only approximate the carbon features with a polynomial fit.
        }
        \label{fig:spec}
    \end{figure*}

    We determine the line-of-sight velocities of our spectra using spexxy\footnote{Available from \url{https://github.com/thusser/spexxy}.}~\citep{Husser-2012-3DSDSP-UG-0}.
    As presented for example by \citet{Husser-2016-A&A-588-A148} for the globular cluster NGC~6397, or for the nearby galaxy NGC~300 by \citet{Roth-2018-A&A-618-A3}, this tool performs a full-spectrum fit to the observed spectrum, using interpolation over a grid of PHOENIX model spectra (the Göttingen Spectral Library) with parameters effective temperature, surface gravity, iron abundance, and alpha-element abundance.
    We fix the alpha-element abundance ratio to solar for our fits, because the quality of the spectra is not high enough to distinguish between different values of this parameter.
    Of the 182~spectra, spexxy is able to fit stellar parameters and line-of-sight velocities for 66 of them.
    For the other 116~spectra the fit failed to converge, typically because of low S/N.
    Fig.~\ref{fig:spec} shows a few example spectra with their spexxy fits.

    Not all sources in our field of view are stars.
    Because we only compare our spectra to stellar templates, we cannot identify and reject non-stars in an automated way.
    In order to minimize contamination by galaxies with a passable stellar fit or by blends between stars and galaxies possibly leading to misidentification of spectral lines, we investigate all extracted spectra with MARZ~\citep{Hinton-2016-A&C-15-61}.
    If a spectrum is clearly galactic or clearly contains galactic lines, or if the source clearly looks like a galaxy in the \emph{HST} image, we exclude the source from our sample.
    This leaves us with 132~spectra that we consider stellar, out of the 182 extracted spectra.
    Fig.~\ref{fig:mag} shows the magnitude distribution of sources meeting this criterion, as well as for additional criteria that we define below.

    For spectra with low S/N spexxy has been found to give unrealistic uncertainty estimates on velocities~\citep{Kamann-2018-MNRAS-473-5591} and we therefore limit our analysis to spectra with $\mathrm{S/N} > 5$, where the noise in the spectrum is estimated from the standard deviation of the spectrum in a $12.5$-$\mathrm{\AA}$ window after removal of a boxcar-smoothed continuum.
    Out of the 182 extracted spectra, 49 satisfy this criterion, while 36 of those are also considered stellar.
    Spexxy was able to determine a velocity for 34 of these spectra.

    Two stars in our field of view have a parallax published in GAIA Data Release~2, one of which is larger than zero at $3\sigma$.
    This parallax places the star at a few kiloparsecs distance from us, so this is clearly not a member of Eri~2 or its putative cluster.
    Its radial velocity is consistent with this, being very significantly different from the systemic velocity of Eri~2.
    The other 180 extracted spectra have no parallax measurement at all and can therefore not be excluded on the basis of distance.
    Taking into account the stellarity, S/N, and velocity criteria as well, we have 33~spectra that satisfy all criteria so far.

\begin{figure*}
        \includegraphics[width=\linewidth]{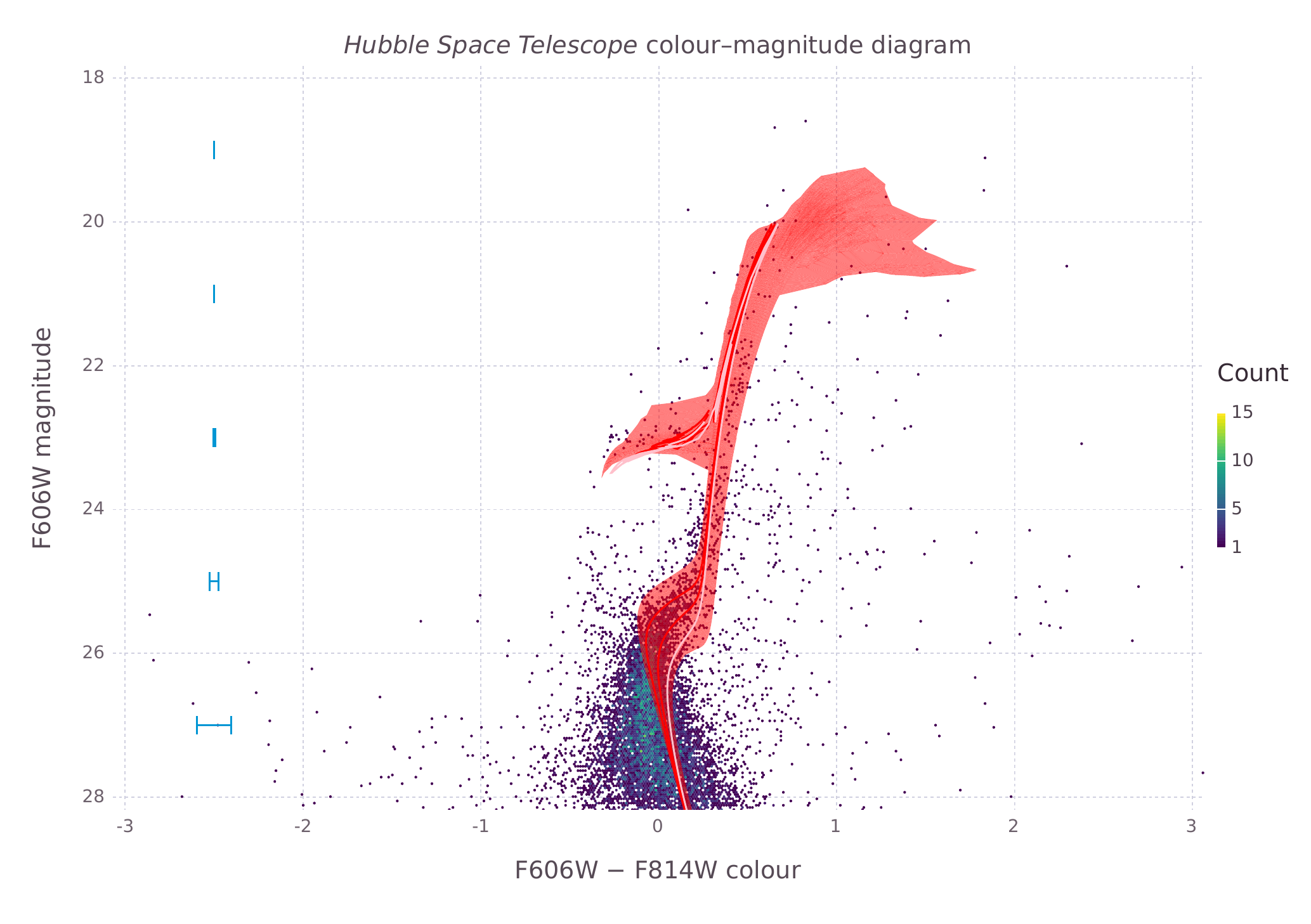}%
        \caption{%
            Colour--magnitude diagram of Eridanus~2 based on photometry of public \emph{Hubble Space Telescope} data.
            Median errors in 2-$\mathrm{mag}$ ranges are indicated with the error bars on the left.
            The red lines are the extremes of the fitting isochrones from the MESA Isochrones \& Stellar Tracks project, with ages of $7$ and $9\,\mathrm{Gyr}$ and a metallicity of $-2.4\,\mathrm{dex}$.
            The red shaded area shows the broadened isochrone obtained by varying the metallicity within its 3-$\sigma$ confidence interval.
            The pink line is a 12-$\mathrm{Gyr}$ isochrone, which is the age estimated by \citet{Crnojevic-2016-ApJL-824-L14}.
            The axis ranges have been chosen to show the majority of the data in greater detail; some data fall outside these ranges and are therefore not visible.
        }
        \label{fig:CMD}
    \end{figure*}

\begin{figure}
        \includegraphics[width=\linewidth]{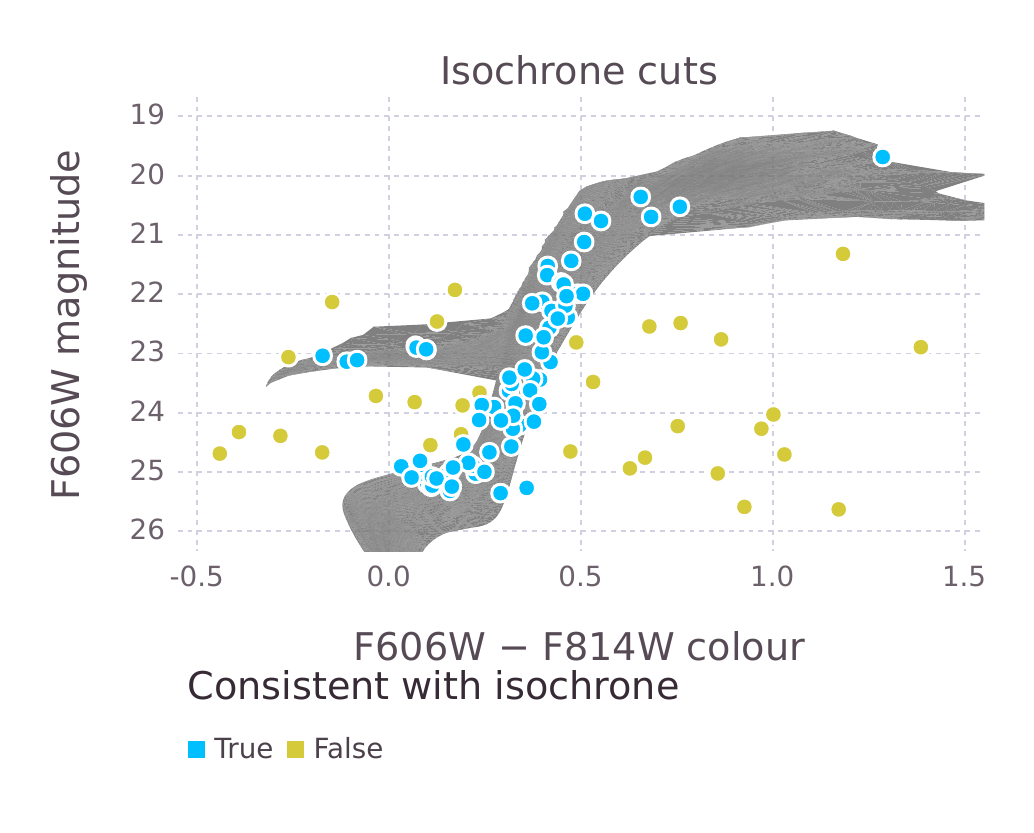}%
        \caption{%
            Selection of Eridanus~2 member stars based on photometric cuts with a broadened isochrone.
            The broadened isochrone has been constructed from 7--9-$\mathrm{Gyr}$ isochrones from the MESA Isochrones \& Stellar Tracks project with a metallicity spread based on previous observations.
            Stars are rejected if they are inconsistent with the broadened isochrone at a 3-$\sigma$ level according to their photometric measurements and uncertainties.
            Two stars with $\mathrm{F606W} - \mathrm{F814W} \approx 2.0$ and $\mathrm{F606W} \approx 25$ are not shown in order to facilitate a more detailed view of the colour-magnitude space around the broadened isochrone.
        }
        \label{fig:isocut}
    \end{figure}

\begin{figure*}
        \includegraphics[width=1.0\linewidth]{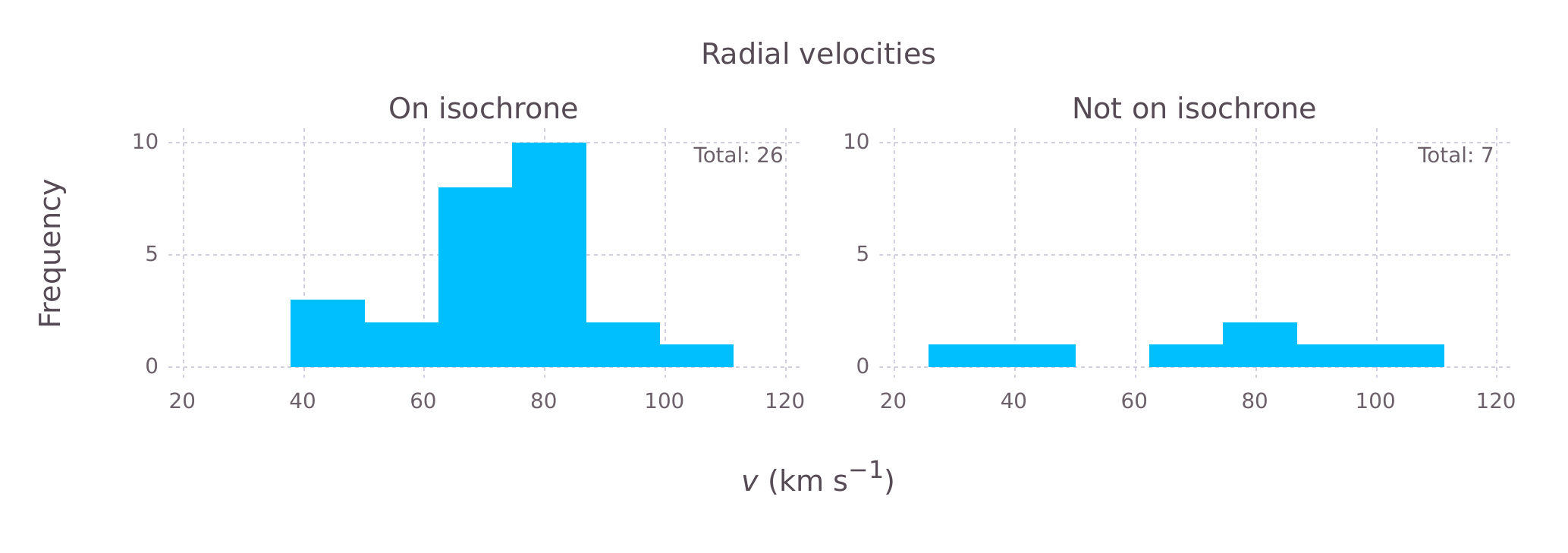}
        \caption{%
            Radial velocities of stars that pass (left panel) or fail (right panel) the isochrone criterion.
            The stars in both panels pass our criteria on stellarity, signal-to-noise ratio, presence of radial velocities, and parallax.
            The mostly flat distribution in the right panel suggests that this criterion succeeds in selecting foreground stars and no large numbers of stars in Eridanus~2.
        }
        \label{fig:isocut_hist}
    \end{figure*}

    Since our observations go deeper than the GAIA data, it is very well possible more foreground stars are present in the remaining selection.
    To further reduce contamination, we perform an isochrone cut.
    First, we match isochrones from the MESA Isochrones \& Stellar Tracks~(MIST) project~\citep{Dotter-2016-ApJS-222-8, Choi-2016-ApJ-823-102, Paxton-2011-ApJS-192-3, Paxton-2013-ApJS-208-4, Paxton-2015-ApJS-220-15} of various ages and with a metallicity of $-2.4\,\mathrm{dex}$ relative to solar --~approximately equal to the best-fit value determined through earlier spectroscopy~\citep{Li-2017-ApJ-838-8}~-- and a V-band attenuation of $0.028\,\mathrm{mag}$~\citep{Schlegel-1998-ApJ-500-525, Schlafly-2011-ApJ-737-103} to the photometry of the public \emph{HST}/ACS data, using the F606W and F814W bands.
    The isochrone ages that give the best match by eye to the data are $7$--$9\,\mathrm{Gyr}$, shown in Fig.~\ref{fig:CMD}.
    This is significantly lower than the $12\,\mathrm{Gyr}$ estimated by earlier photometry~\citep{Crnojevic-2016-ApJL-824-L14}.
    Because \citet{Li-2017-ApJ-838-8} found a significant spread in metallicity, we use 7-, 8-, and 9-Gyr isochrones with metallicities between $-3.8\,\mathrm{dex}$ and $-1.0\,\mathrm{dex}$, approximately corresponding to three times the dispersion around the mean value, in steps of $0.1\,\mathrm{dex}$.
    This seems to capture the main-sequence turn-off quite well by eye.
    To turn this collection of isochrones into a selection region in colour--magnitude space, we construct a broadened isochrone using an $\alpha$-shape \citep{Edelsbrunner-1983-ITIT-29-551}, shown in Fig.~\ref{fig:CMD} as the red shaded area.
    We then draw 5000 sample photometric measurements from a normal distribution for each star for which we have both a spectrum and two-colour photometry, with mean and standard deviation given by the measured flux and uncertainty on flux.
    We used a conservative rejection criterion to avoid the exclusion of real cluster member stars, rejecting stars from further analysis if they fall outside of the broadened isochrone at a 3-$\sigma$ confidence level over the 5000 draws.
    Of the 113~stars for which we have both a spectrum and two-colour photometry, 79~stars are not rejected based on this criterion (see Fig.~\ref{fig:isocut}).
    Twenty-six of these stars also satisfy the criteria mentioned earlier in this subsection.
    When we compare the radial velocities of the stars that do and do not match the isochrones (see Fig.~\ref{fig:isocut_hist}), we see a peaked distribution of accepted stars and a mostly flat distribution of rejected stars.
    This is what we expect for Eri~2 and the Milky Way, respectively, though we do see a flat background in the accepted stars that could be caused by uncaught Milky Way stars.

    Additionally, we try to further constrain contamination using the measured velocities, though ultimately this does not reject any extra stars.
    Based on the Besan\c{c}on model of the Milky Way~\citep{Robin-2003-A&A-409-523, Robin-2004-A&A-416-157}, we expect remaining foreground Milky-Way stars to have a velocity distribution that can be approximated with a single linear slope between $v_\mathrm{min} \coloneqq 10\,\mathrm{km}\,\mathrm{s}^{-1}$ and $v_\mathrm{max} \coloneqq 120\,\mathrm{km}\,\mathrm{s}^{-1}$, while members of Eri~2 are expected to behave approximately according to a normal distribution.
    The distribution of Milky Way--star velocities is sufficiently broad that we can approximate members and non-members of Eri~2's cluster, if it exists, with a single distribution.
    We use a maximum-likelihood approach \citep[see e.g.][]{Hargreaves-1994-MNRAS-269-957, Martin-2018-ApJL-859-L5}.
    Contrary to \citet{Martin-2018-ApJL-859-L5}, we consider a membership probability for each star instead of a global contamination fraction.
    The likelihood is then
    \begin{equation}
        \begin{split}
            \mathcal{L}(\mu_\mathrm{int}, \sigma_\mathrm{int}, m_i | v_i, \varepsilon_i)
            &= \prod_i \Bigg[\frac{m_i}{\sqrt{2\pi}\sigma_{\mathrm{obs},i}} \exp \Bigg(-\frac{1}{2} \bigg(\frac{v_i-\mu_\mathrm{int}}{\sigma_{\mathrm{obs},i}}\bigg)^2\Bigg)\\
            &\qquad\quad+ (1-m_i)(av_i + b)\Bigg],
        \end{split}
    \end{equation}
    where $\mu_\mathrm{int}$ and $\sigma_\mathrm{int}$ are the intrinsic mean velocity and intrinsic velocity dispersion of Eri~2, $\sigma_{\mathrm{obs},i} = (\sigma_\mathrm{int}^2 + \varepsilon_i^2)^{1/2}$ is the observed velocity dispersion for star~$i$, $m_i$ is the probability for star~$i$ to be a member of Eri~2, $v_i$ and $\varepsilon_i$ are the measured velocity and velocity uncertainty of star~$i$, $a$ is the slope of the approximated Milky Way--star distribution, and $b$ is a constant to normalize that distribution, defined as
    \begin{equation}
        b \coloneqq \frac{1-\frac{a}{2}(v_\mathrm{max}^2-v_\mathrm{min}^2)}{v_\mathrm{max}-v_\mathrm{min}}.
    \end{equation}
    Using a maximum-likelihood optimization, we find a slope $a \approx -4.3 \times 10^{-5}\,\mathrm{s}^2\,\mathrm{km}^{-2}$.

    To determine an accurate value of $\sigma_\mathrm{int}$, it is crucial to have properly estimated velocity uncertainties~$\varepsilon_i$.
    \citet{Kamann-2016-A&A-588-A149} showed that the uncertainties estimated by spexxy are an accurate description of the true uncertainties, even without further calibration, which would require more spectra and more epochs than we have.
    For the same reason, we cannot correct for or exclude binary stars in our sample with the current data.

    We compute the above likelihood function on a 101-by-301-by-301 linearly spaced grid for $m_i = 0$--$1$, $\mu_\mathrm{int} = 60$--$90\, \mathrm{km}\, \mathrm{s}^{-1}$, and $\sigma_\mathrm{int} = 0$--$30\, \mathrm{km}\, \mathrm{s}^{-1}$, using uniform priors.
    The different $m_i$ can be marginalized separately for each~$i$, over $\mu_\mathrm{int}$ and $\sigma_\mathrm{int}$.
    The values of $m_i$ depend on the choice of $v_\mathrm{min}$ and $v_\mathrm{max}$ and are therefore not very meaningful in the absolute sense, but comparing them to each other could reveal outliers.
    We see no evidence of this; the estimated membership probabilities are all similar.
    We also performed the same procedure with a uniform instead of a sloped distribution.
    This gave slightly different membership probabilities, but lead to the same conclusion.
    Our final selection therefore remains the same, containing 26~stars.

    We identify by eye a circular region in our white-light image and in the \emph{HST}/ACS images at MUSE coordinates $3\mathrm{h}44\mathrm{m}22\fs3$, $-43\degr31\arcmin58\farcs8$ and radius $12.5\,\mathrm{arcsec}$ that we consider the projected area of the putative cluster.
    This area is indicated with a green circle in the left panel of Fig.~\ref{fig:im}.
    Seven of the finally selected stars are inside this region.

\begin{figure}
        \includegraphics[width=1.0\linewidth]{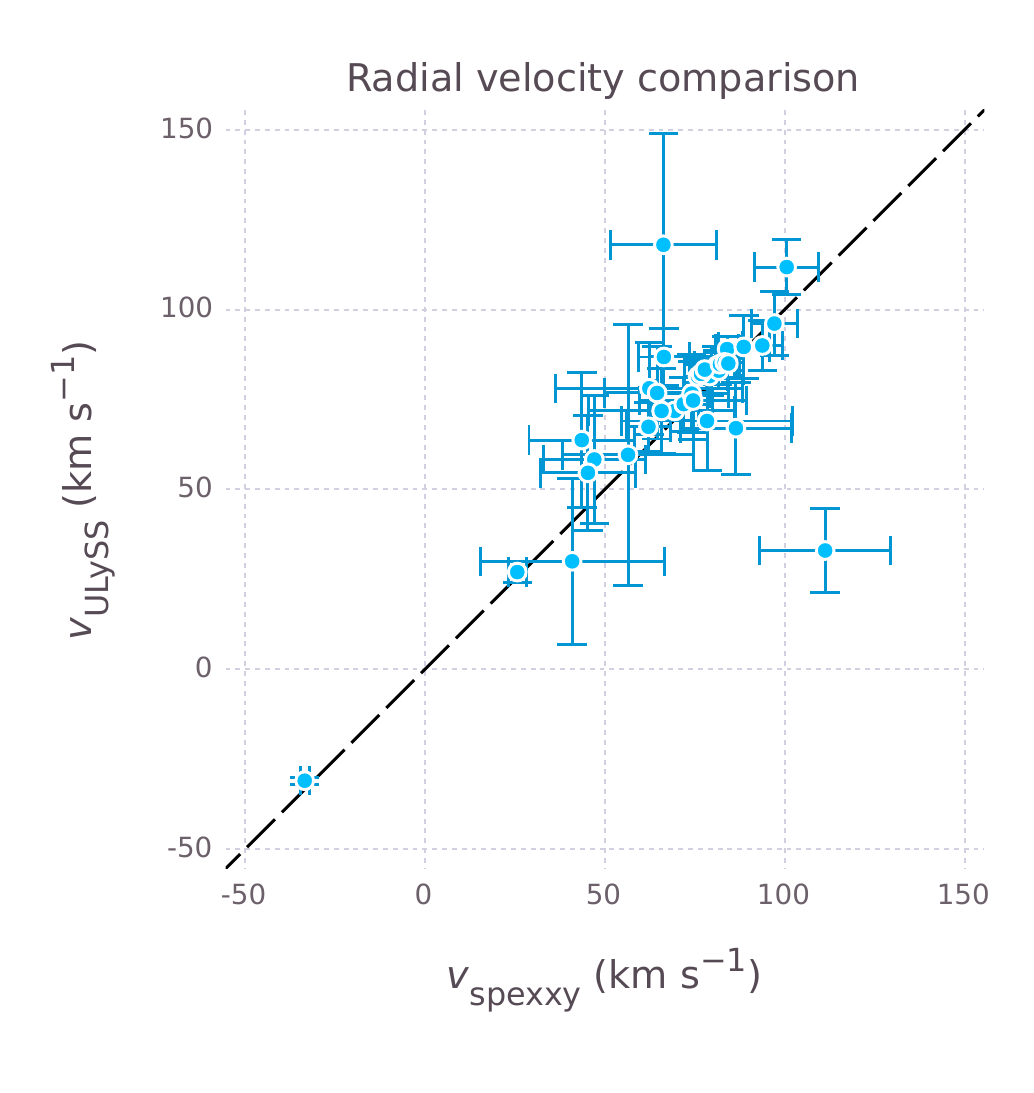}%
        \caption{%
            Comparison of line-of-sight velocity estimates from the full-spectrum fitting codes spexxy and ULySS.
            The estimates from spexxy are obtained by fitting with an interpolation over the PHOENIX library of synthetic spectra, while ULySS uses a linear combination of empirical spectra from the MIUSCAT library.
            As the ULySS uncertainties were overpredicted for higher signal-to-noise sources, an empirical correction has been applied.
            The two methods show reasonable agreement, save for a small systematic shift towards higher velocities for ULySS.
        }
        \label{fig:uvss}
    \end{figure}

    As an independent check, and following the approach for NGC~300 \citep{Roth-2018-A&A-618-A3}, we use the ULySS code and the empirical MIUSCAT library as an alternative to fit the spectra.
    Of the 182~spectra, ULySS is able to provide 95~fits for a subsample with a S/N cut of $2$.
    The fits measure line-of-sight velocities, and yield values for the effective temperature, gravity, and metallicity of the best fitting template star spectra from the library.
    From the latter, we are able to constrain the spectral type classification of the objects.
    By visual inspection, 19 objects are immediately rejected as background galaxies, and two main-sequence star spectra are rejected as foreground stars.
    The same stars were also rejected in our standard analysis above.
    On the basis of a Monte-Carlo simulation with Poissonian noise applied to seven selected template star spectra, involving 900~realizations with S/N values between $1$ and $30$, we discover that the formal line-of-sight velocity errors provided by ULySS are overestimates below a S/N level of $20$.
    We therefore apply an empirical correction to the velocity uncertainties.
    The simulation also indicates that the velocities for hot stars may be overestimated.
    Fig.~\ref{fig:uvss} shows that the velocity measurements obtained by spexxy and ULySS for the sample of 34~sources that have stellar spectra with a S/N above $5$ and a spexxy fit are in reasonable agreement between each other, except for a small systematic shift.
    This shift may be due to the overestimation of radial velocities for blue stars, and the general effect where blending of different empirical templates at low S/N can lead to confusion of different spectral lines.
    As the synthetic PHOENIX spectra and single-template fitting by spexxy are not affected by these issues, we will continue to use the spexxy velocity estimates, though the templates identified by ULySS still give a good idea of the spectral types.

\begin{figure*}
        \includegraphics[width=1.0\linewidth]{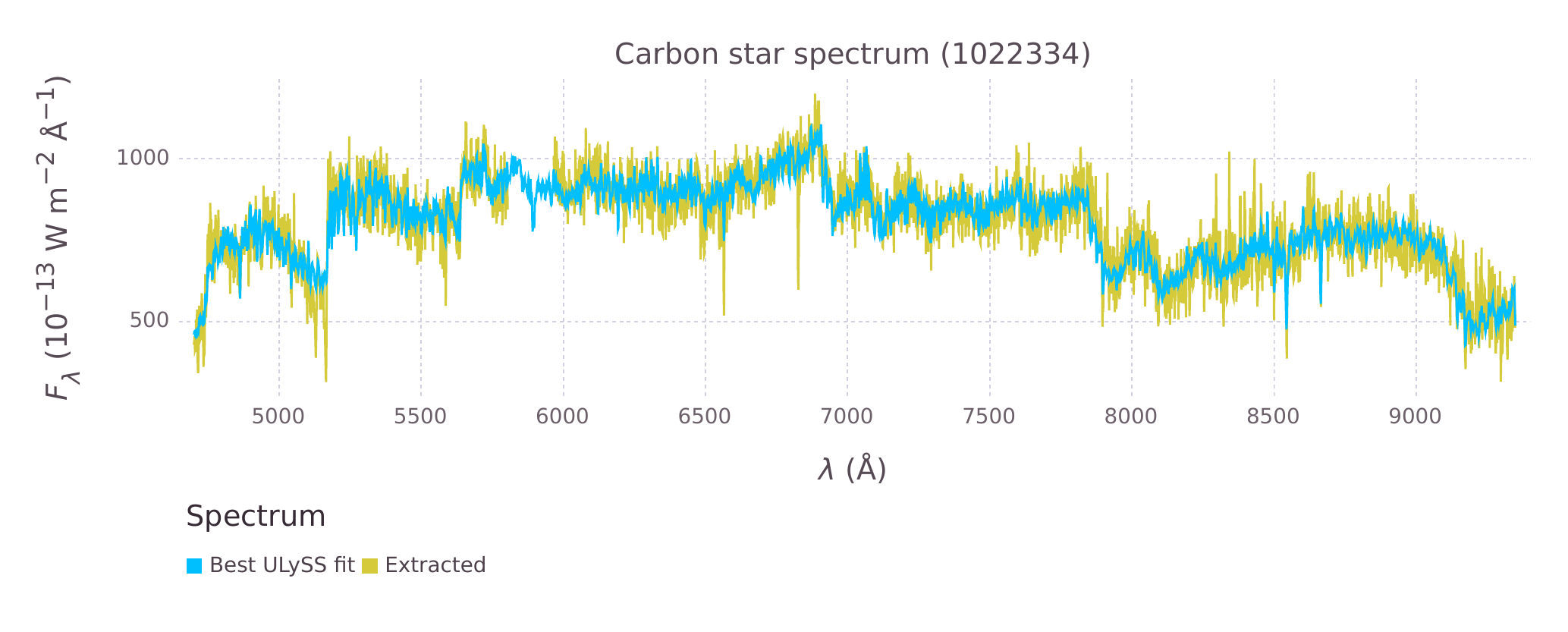}
        \caption{%
            The extracted spectrum of carbon star 1022334 (same as in Fig.~\ref{fig:spec}, bottom panel) in yellow, shown with the best fit determined with ULySS in blue, using the empirical MIUSCAT spectral library amended with 19~spectra of carbon stars from the XShooter library.
            Contrary to the synthetic fit, this fit properly reproduces the CN features, including the sharp band heads, with the exception of the one at ${\sim}5150\, \mathrm{\AA}$.
            What may seem to be noise in the fitted spectra is in fact a forest of molecular lines unresolved at the resolution of MUSE.
        }
        \label{fig:cstar}
    \end{figure*}

    Visually inspecting the spectra, we find there is one notable exception for the agreement between the fits, namely the spectrum for the star with ID~1022334, listed with a S/N estimate of 15 (see Fig.~\ref{fig:spec}, bottom panel).
    It shows a zig-zag appearance of the continuum, and a prominent discontinuity at a wavelength near $5165\,\AA$, such that neither spexxy nor ULySS are able to fit the continuum, except the absorption lines of the calcium triplet and the Balmer lines of $\mathrm{H\alpha}$ and $\mathrm{H\beta}$.
    Comparison with the XShooter library spectra and models presented by \citet{Gonneau-2016-A&A-589-A36, Gonneau-2017-A&A-601-A141}, we discover that this star clearly shows the $\mathrm{C}_2$ and $\mathrm{CN}$ molecular bands that are characteristic for a carbon star.
    On the basis of photometry and the measured line-of-sight velocity of $71 \pm 2\,\mathrm{km}\,\mathrm{s}^{-1}$, we conclude that this star is indeed a carbon star and a member of Eridanus 2.
    Prompted by this discovery, we amend the MIUSCAT library with 19~spectra from the XShooter library\footnote{Available from \url{http://xsl.u-strasbg.fr/index_files/carbon}.} and repeat the ULySS fitting procedure.
    As a result, we find a total of three carbon star candidates, with IDs 1022334 (excellent fit, S/N of 15, see also Fig.~\ref{fig:cstar}), 1016071 (likely, S/N of 9), and 11724 (possible, S/N of 4).
    The kinematics of all three stars are consistent with membership of Eri~2, but only the star with ID~1022334 meets our strict membership criteria.
    The other two stars are excluded from our analysis because of a S/N below $5$ (ID~11724) or photometry\footnote{Due to the characteristic spectral features of carbon stars, photometric classification can be misleading, but to avoid possible foreground contamination we do not relax our criteria for membership.} that does not match the Eri~2 isochrones (ID~1016071), but membership of Eri~2 cannot be ruled out.
    Due to their locations on the sky, the star with ID~1016071 is possibly associated with the Eri~2 cluster, while this cannot be for the other two.
    As carbon stars are thermally-pulsating asymptotic giant--branch stars, these objects must belong to an intermediate-age stellar population, which would be a spectroscopic confirmation of the younger population in Eridanus~2 suggested by \citet{Koposov-2015-ApJ-805-130}.

    \refstepcounter{table}
    \label{tab:sourcelist}
    We provide a list of 35~sources that have a S/N of at least $5$, a spectrum that we consider stellar, and a measurement of the line-of-sight velocity by spexxy; or spectral features characteristic for a carbon star; in Table~\ref{tab:sourcelist}, available at the CDS, which contains the following information: Column~1 lists the source~ID, Columns~2 and~3 give the right ascension respectively declination calibrated to GAIA Data Release~2, Column~4 gives the angular separation from the centre of the cluster as we defined it, Column~5 lists the S/N, Columns~6 and~7 give the line-of-sight velocity and associated error determined using spexxy, Columns~8 and~9 give the metallicity and associated error determined using spexxy (presented in Sect.~\ref{ssec:properties}), Columns~10 and~11 list the parallax and associated error from GAIA Data Release~2, Columns~12 and~13 give the apparent magnitudes from the F606W respectively F814W public \emph{HST} data, Column~14 gives the likelihood of that photometry being consistent with the broadened isochrone, Column~15 indicates whether the source is considered a member of Eri~2, Column~16 gives the membership likelihood for possible cluster member stars (presented in Sect.~\ref{ssec:distributions}), Columns~17 and~18 give the line-of-sight velocity and associated error determined using ULySS, Column~19 gives the spectral type determined using ULySS, and Column~20 gives the metallicity determined using The Cannon (presented in Sect.~\ref{ssec:properties}).

\subsection{Velocity distributions}
\label{ssec:distributions}
    After the effort to remove foreground stars and background objects, we assume all remaining stars are part of Eri~2 or its cluster.
    We assume the line-of-sight velocities are intrinsically normally distributed and that the dwarf galaxy and putative cluster may have different distributions.
    To determine the parameters, using the same definitions as in Sect.~\ref{ssec:velocities}, we calculate the likelihood
    \begin{equation}
        \mathcal{L}(\mu_\mathrm{int}, \sigma_\mathrm{int} | v_i, \varepsilon_i) = \prod_i \Bigg[\frac{1}{\sqrt{2\pi}\sigma_{\mathrm{obs},i}} \exp \Bigg(-\frac{1}{2} \bigg(\frac{v_i-\mu_\mathrm{int}}{\sigma_{\mathrm{obs},i}}\bigg)^2\Bigg)\Bigg],
    \end{equation}
    separately for the cluster and the bulk of Eri~2.

    A complicating factor is that we do not know which stars are members of the putative star cluster.
    We see seven stars in the projected area of the cluster, but that does not necessarily imply membership.
    Though it is possible to use a Markov-chain Monte-Carlo approach with membership probabilities for these seven stars, the low dimensionality of this problem makes it possible to perform a deterministic calculation.
    With seven potential members of the cluster, we can distinguish $2^7 = 128$ different scenarios.
    For each scenario, a potential cluster member is either included in the cluster or in the bulk.
    The length of $i$ therefore depends on the scenario and ranges from 0 to 7 for the cluster and from 19 to 26 for the bulk of Eri~2.

    For each scenario, the likelihood for the cluster and for the bulk can be calculated separately, because they have independent kinematic distributions.
    We calculated the above likelihoods for each of these scenarios on a 301-by-301 linearly spaced grid for $\mu_\mathrm{int} = 60$--$90\,\mathrm{km}\,\mathrm{s}^{-1}$ and $\sigma_\mathrm{int} = 0$--$30\,\mathrm{km}\,\mathrm{s}^{-1}$, with uniform priors.
    As the total likelihood is the product of the likelihoods for cluster and bulk, we multiply the cluster likelihood grid with the marginalized bulk likelihood and vice versa, for each scenario.

    Each scenario also has a prior likelihood associated with it, based on the expected number of cluster members.
    Comparing the density of stars outside the cluster with the projected area of the cluster, we expect ${\sim}1.82$ non-member stars inside this area.
    The prior membership probability for each star is therefore $p_{\mathrm{mem},\mathrm{pri}} \coloneqq 1 - 1.82/7 \approx 0.740$, and the prior probability for the scenario~$s$ with $n_\mathrm{mem}$ member stars follows from
    \begin{equation}
        p_{s,\mathrm{pri}} = (p_{\mathrm{mem},\mathrm{pri}})^{n_\mathrm{mem}} (1-p_{\mathrm{mem},\mathrm{pri}})^{7-n_\mathrm{mem}}.
    \end{equation}
    We multiply the calculated grids of likelihoods for each scenario with the respective prior factor and marginalize over the membership scenarios.

\begin{figure}
        \includegraphics[width=1.0\linewidth]{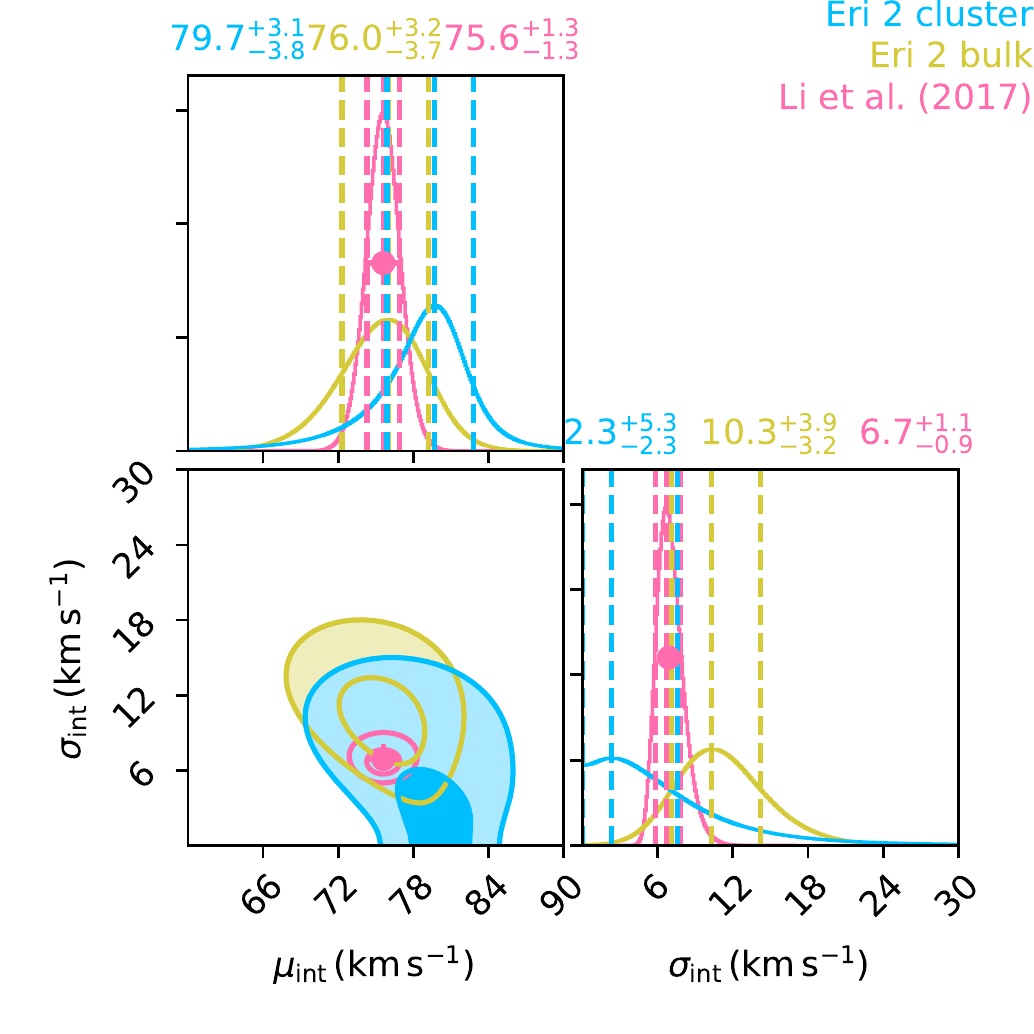}%
        \caption{%
            Corner plot of the distribution of line-of-sight velocities for stars in Eridanus~2, assuming normal distributions with intrinsic mean~$\mu_\mathrm{int}$ and intrinsic dispersion~$\sigma_\mathrm{int}$.
            The blue, yellow, and pink contours indicate the 39- and 86-$\%$ confidence areas (equivalent to 1- and 2-$\sigma$ for a two-dimensional normal distribution) for stars in the putative star cluster (marginalized over cluster membership), stars in the bulk of Eridanus~2 (marginalized over cluster membership for possible cluster member stars), and stars at a larger radius as measured by \citet{Li-2017-ApJ-838-8}.
            The maxima of the one-dimensional posteriors and the 68-$\%$ confidence interval around it are indicated in the same colours above the on-diagonal panels.
            The pink circle and error bars indicate the result \citet{Li-2017-ApJ-838-8} themselves derive from their measurements.
        }
        \label{fig:cornerlike}
    \end{figure}

    Fig.~\ref{fig:cornerlike} shows a corner plot of the final grids for the cluster and the bulk of Eri~2, where we also compare against the distributions found by \citet{Li-2017-ApJ-838-8} and a reconstruction of those distributions using their velocities with our calculation.
    The best-fit values, $\mu_\mathrm{int} = 79.7^{+3.1}_{-3.8}\, \mathrm{km}\, \mathrm{s}^{-1}$ and $\sigma_\mathrm{int} = 2.3^{+5.3}_{-2.3}\, \mathrm{km}\, \mathrm{s}^{-1}$ for the cluster and $\mu_\mathrm{int} = 76.0^{+3.2}_{-3.7}\, \mathrm{km}\, \mathrm{s}^{-1}$ and $\sigma_\mathrm{int} = 10.3^{+3.0}_{-3.2}\, \mathrm{km}\, \mathrm{s}^{-1}$ for the bulk, are determined by the maximum values of the posterior likelihoods, while the confidence intervals are derived by sequentially adding the point with the highest posterior likelihood outside of the confidence interval, until a 68-$\%$ confidence level is reached.
    The velocity dispersion of the cluster is consistent with zero; its 68- and 95-$\%$ upper limits are $7.6\, \mathrm{km}\, \mathrm{s}^{-1}$ and $17.5\, \mathrm{km}\, \mathrm{s}^{-1}$, respectively.

\begin{figure*}
        \includegraphics[width=\linewidth]{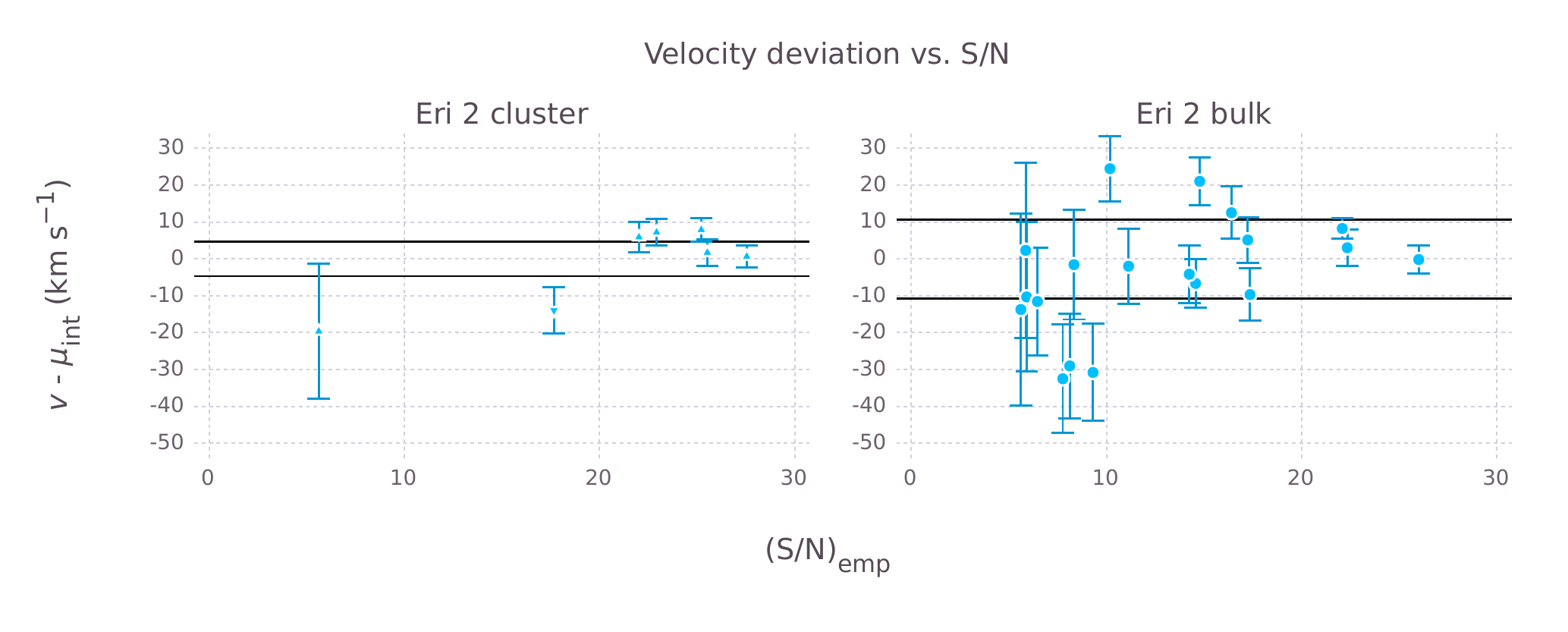}%
        \caption{%
            Line-of-sight velocities and measurement uncertainties for stars in Eridanus~2, relative to the intrinsic mean velocity, as a function of signal-to-noise ratio.
            The possible cluster members are marked with triangles, pointing upward for those with a high membership probability (${\sim}80$--$90\,\%$) and downward for the one source with a lower probability (${\sim}60\,\%$).
            The other stars are marked with circles.
            The black horizontal lines indicate the intrinsic velocity dispersion.
            The most likely scenario, where all possible cluster member stars are indeed cluster members, is shown.
            \emph{Left:}
            Velocities for member stars of the cluster.
            \emph{Right:}
            Velocities for stars in the bulk of Eridanus~2.
        }
        \label{fig:velocities}
    \end{figure*}

    Alternatively we can marginalise over the mean velocities and velocity dispersions to get constraints on the membership of each star.
    For six out of seven, the posterior membership probability is ${\sim}80$--$90\,\%$, indicating that the velocity data of these stars are better described with a separate distribution rather than the same distribution as the bulk of Eri~2.
    The remaining star has a slightly lower membership probability of ${\sim}60\,\%$, which indicates the kinematics disfavour its membership, but overall it still provides a significant contribution to the cluster parameters.
    It is not necessary to exclude it from the analysis, as the combinatorial nature of our calculation takes its lower membership probability (and those of the other potential members) into account.
    We list the membership probabilities in Table~\ref{tab:sourcelist}, available at the CDS.
    We plot the velocities relative to the means of the two populations as a function of S/N in Fig.~\ref{fig:velocities}, indicating high- and low-probability cluster members with different symbols.
    We note that there are two low-S/N stars in the projected area of the cluster that have somewhat lower velocities than the others.
    The measurement uncertainties are large enough, however, to make the measurements consistent with the distribution.
    In the larger sample of non-member stars we can see a clear trend between measurement uncertainty and S/N.
    Here, as well, there are no clear outliers.
    This means that, if there are still contaminants present, they will have little effect on the velocity distributions that we have determined.

\subsection{Properties of Eridanus~2 and its cluster}
\label{ssec:properties}
    There is a small tension between the velocity dispersions of the bulk of Eri~2 and the putative star cluster.
    This is a point in favour of the latter being a dynamically different population of stars.
    The lower value of the velocity dispersion is consistent with a stellar cluster as opposed to a galaxy.
    Finally, the agreement in mean velocity suggests the two populations are related and that the star cluster is either in or orbiting Eri~2, as opposed to being a foreground or background object.
    If we select seven random stars as possible cluster members instead of the seven in the projected area of the cluster, we typically retrieve two distributions that are completely consistent with each other, both with a non-zero velocity dispersion.
    This is also in line with the existence of a cluster.

    This spectroscopic evidence enhances the claim based on photometry that there is a star cluster inside Eri~2.
    For the remainder of this article, we will therefore assume the cluster is a real, dynamically distinct stellar structure in Eri~2 and that we can therefore use our measurements to constrain MACHOs as dark matter.

    Based on the measured velocity dispersion~$\sigma_\mathrm{cl}$, we can estimate the dynamical half-light mass of the star cluster using the estimator by \citet{Wolf-2010-MNRAS-406-1220}:
    \begin{equation}
        M_{\mathrm{cl},\mathrm{dyn}} = \frac{4\sigma_\mathrm{cl}^2R_{\mathrm{h},\mathrm{cl}}}{G}.
    \end{equation}
    If we perform this calculation on the posterior of $\sigma_\mathrm{cl}$, using $R_{\mathrm{h},\mathrm{cl}} = 13\,\mathrm{pc}$~\citep{Crnojevic-2016-ApJL-824-L14}, we obtain $M_{\mathrm{cl},\mathrm{dyn}} = 6.4^{+63.4}_{-6.4} \times 10^4\,M_\sun$, which due to the uncertainty on the velocity dispersion is consistent with zero, with 68- and 95-$\%$ upper limits of $7.0 \times 10^5\,M_\sun$ and $3.70 \times 10^6\,M_\sun$, respectively.

    The dynamical mass estimate includes a, likely dominant, contribution by the background dark-matter distribution of Eri~2.
    Another option to constrain the cluster's mass is to use its luminosity and adopt a mass-to-light ratio, which should give an estimate of the baryonic mass.
    We determine mass-to-light ratios for several initial mass functions~(IMFs) using the various MIST isochrones described in Sect.~\ref{ssec:velocities}.
    To begin with, we consider the well-known \citet{Salpeter-1955-ApJ-121-161} and \citet{Kroupa-2001-MNRAS-322-231} IMFs.
    The Salpeter IMF is a single power law $\mathrm{d}N/\mathrm{d}M \propto M^{-\alpha}$ with slope $\alpha = -2.35$, while the Kroupa IMF is a broken power law with $\alpha = 0.3$ for $M < 0.08\,M_\sun$, $1.3$ for $0.08\,M_\sun \le M < 0.5\,M_\sun$, and $2.3$ for $M \ge 0.5\,M_\sun$.
    Observations, on the other hand, are at slight tension with these IMFs and show slopes of ${\sim}1.2$ between roughly $0.5$--$0.8\,M_\sun$~\citep{Geha-2013-ApJ-771-29}.
    For completeness, we therefore also consider modified versions of the Salpeter and Kroupa IMFs, where we change the slope of the Salpeter IMF from $2.35$ to $1.2$ and extend the middle segment of the Kroupa IMF with slope $1.3$ up to $0.8\,M_\sun$ instead of $0.5\,M_\sun$.
    In the calculation of the mass-to-light ratio, we include the masses of stellar remnants according to \citet{Renzini-1993-ApJ-416-L49}, as these stellar remnants will also participate in the dynamical interactions with MACHOs.
    The standard Salpeter and Kroupa IMFs and the modified Kroupa IMF give comparable results, with typically a ${\sim}20\,\mathrm{\%}$ variation over the range of metallicities.
    The modified Salpeter IMF gives ${\sim}10$ times higher results, because of the relatively large number of massive stellar remnants due to the shallow single slope.
    We will assume the standard Kroupa IMF at a metallicity of $-2.4\,\mathrm{dex}$ as our fiducial IMF and will explore the impact of using different IMFs at a later point.
    The mass-to-light ratio for this IMF is ${\sim}1.43$ for a 7-$\mathrm{Gyr}$-old population, ${\sim}1.56$ for $8\,\mathrm{Gyr}$, and ${\sim}1.67$ for $9\,\mathrm{Gyr}$, giving a total cluster mass of $M_\mathrm{cl} = 3.0 \times 10^3\,M_\sun$, $3.3 \times 10^3\,M_\sun$, and $3.5 \times 10^3\,M_\sun$, respectively, using the cluster's total absolute V-band magnitude $M_\mathrm{V} = -3.5$ calculated by \citet{Crnojevic-2016-ApJL-824-L14}.
    The IMF-based estimates are consistent with being smaller than the dynamical estimate, as expected.
    When considering the mass of the star cluster itself, without the dark-matter background, we will use the IMF-based estimates.

    With the current data it is difficult to make any statement on the density profile of Eri~2.
    \citet{Contenta-2018-MNRAS-476-3124} argue that, if the cluster is located near the centre, Eri~2 needs to have a cored density profile, otherwise the tidal forces would visibly disturb the cluster.
    On the other hand, the probability of a chance alignment between a cluster far away from a cuspy centre is so low that the first scenario is strongly preferred.
    Like \citet{Brandt-2016-ApJL-824-L31}, we will therefore assume the cluster is located inside a central core.

\begin{figure*}
        \includegraphics[width=\linewidth]{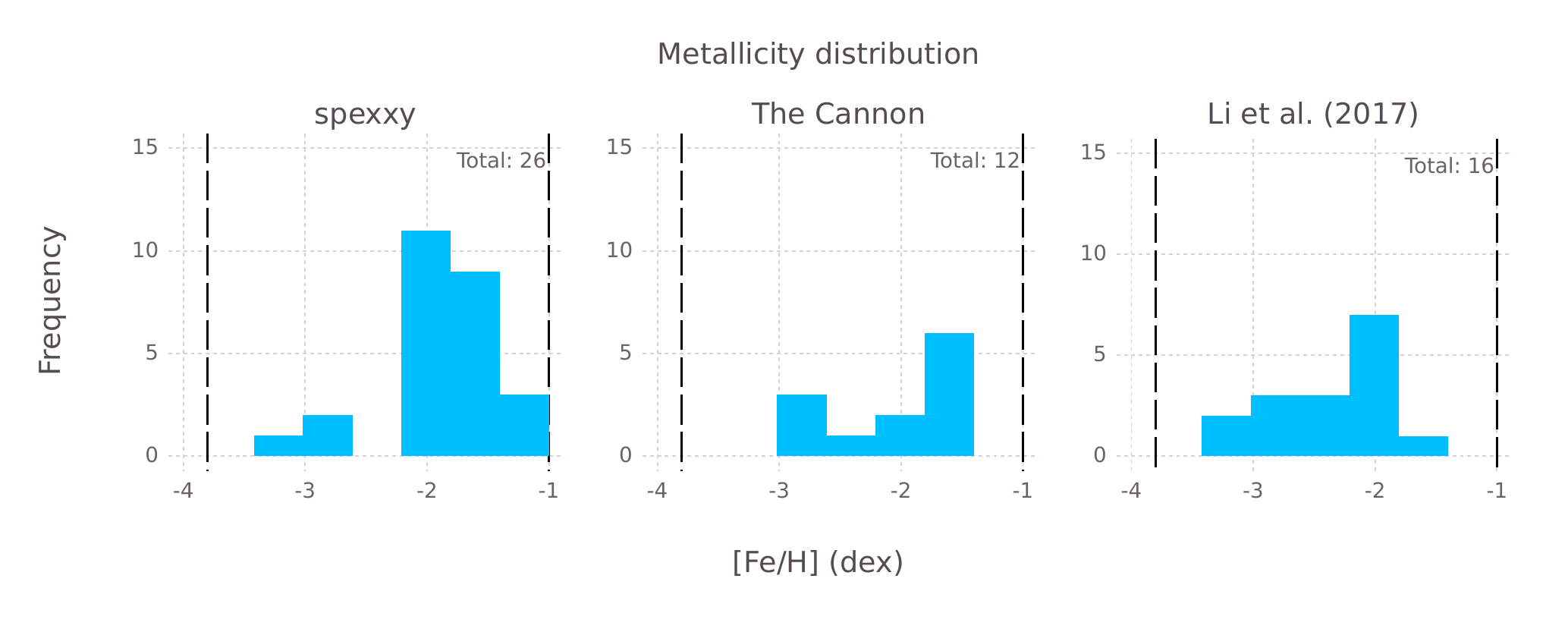}
        \caption{%
            Metallicity distributions of stars in Eridanus~2.
            From left to right, metallicities as predicted by full-spectrum fitting with a synthetic stellar-template library, predicted metallicities based on our observations for stars with a signal-to-noise ratio $\mathrm{S/N} > 10$ determined with a transfer-learning algorithm, and metallicities by \citet{Li-2017-ApJ-838-8} for a different sample of stars.
            The black vertical dashed lines indicate the range of metallicities we have used for isochrone calculation, based on the distribution determined by \citet{Li-2017-ApJ-838-8}.
        }
        \label{fig:FeH}
    \end{figure*}

    We determine metallicities for our sample in two different ways: with spexxy and with The Cannon~\citep{Ness-2015-ApJ-808-16}, a transfer-learning algorithm that predicts stellar parameters from spectra.
    The Cannon is trained on a set of stars from the MUSE Survey of Galactic Globular Clusters~\citep{Kamann-2018-MNRAS-473-5591} to create a model that enables us to predict the metallicity.
    We use 15~globular clusters from the survey and select literature values from works with large samples of stars with estimates of stellar parameters.
    We select stars in a temperature range of $3500$--$7000\,\mathrm{K}$, with defined radial velocities and similar magnitudes as the ones observed with MUSE.
    These requirements lead us to a training set of 176~stars with 2409~spectra.
    We normalize the continuum of the spectra and re-bin the data to have the same number of bins for the training and test sets.
    We also censor the wavelengths where the telluric lines can bias the estimations.
    The trained model is good enough for our purpose with a root-mean-square of $0.16$ and an $r^2$-score of $0.89$.
    We apply this model on Eri~2 stars with a $\mathrm{S/N} > 10$.
    In Fig.~\ref{fig:FeH} we present the metallicity distributions given by these two methods and compare to the results of \citet{Li-2017-ApJ-838-8} as well.
    Spexxy and The Cannon are in reasonable agreement, while the literature sample seems to have a slightly lower metallicity on average.
    In all cases, the metallicities are well within the range of metallicities used in the isochrone cut.
    The metallicity measurements for 35~sources with S/N above $5$, a spectrum that we consider stellar, and a measurement of line-of-sight velocity using spexxy; or spectral features characteristic for carbon stars; are provided in Table~\ref{tab:sourcelist}, available at the CDS.

\section{Implications for MACHOs}
\label{sec:impl}
    Now that we have determined the cluster in Eri~2 likely exists, we can use the measured velocity dispersions to put constraints on MACHOs.
    First we describe a mathematical model for the effect the MACHOs have on the cluster (Sect.~\ref{ssec:model}).
    We then use this to constrain the contribution of MACHOs to dark matter as a function of their mass (Sect.~\ref{ssec:constraints}).

\subsection{Mathematical model for the disruption of the Eridanus~2 cluster}
\label{ssec:model}
    To derive constraints on MACHOs from our observations, we mostly follow the same derivation as \citet{Brandt-2016-ApJL-824-L31} and derive a limiting MACHO fraction as a function of MACHO mass and observed velocity dispersions.
    We deviate at some points from \citeauthor{Brandt-2016-ApJL-824-L31}, which are detailed in this subsection.
    The derivation itself is detailed in Appendix~\ref{app:fMlim}.
    It is based on the assumption that the stars in the cluster and the MACHOs are two distinct dynamical populations and that MACHOs that pass through the star cluster transfer energy to the stars through dynamical interactions.
    The energy transfer is calculated with a diffusion approximation and results in a growing cluster half-light radius, leading eventually to dissipation of the cluster.
    By comparing the dissipation timescale with the cluster's age, we can constrain the MACHO abundance and mass that enter into the diffusion calculation.

    The dark matter is assumed to consist for a fraction~$f_\mathrm{M}$ of MACHOs and for the remaining fraction of cold, collisionless particles.
    Contrary to \citet{Brandt-2016-ApJL-824-L31}, we maintain the diffusion coefficients for cooling from the cold, collisionless dark matter.
    We also use a different measure for the dissipation timescale, namely $R_{\mathrm{h},\mathrm{cl}}/\dot{R}_{\mathrm{h},\mathrm{cl}}$.
    This leads us to a limiting MACHO fraction
    \begin{equation}
        f_{\mathrm{M},\mathrm{lim}} \coloneqq \frac{\dot{U}_{\mathrm{cl},\mathrm{lim}} - 2\mathrm{D}[\Delta E_\mathrm{cl}]\Big|_\mathrm{C}}{2\mathrm{D}[\Delta E_\mathrm{cl}]\Big|_\mathrm{M} - 2\mathrm{D}[\Delta E_\mathrm{cl}]\Big|_\mathrm{C}},
    \end{equation}
    where
    \begin{equation}
        \dot{U}_{\mathrm{cl},\mathrm{lim}} \coloneqq \frac{\alpha GM_\mathrm{cl}^2 + 2\beta G\rho M_\mathrm{cl}R_{\mathrm{h},\mathrm{cl}}^3}{R_{\mathrm{h},\mathrm{cl}}t_\mathrm{cl}}
    \end{equation}
    describes the increase in potential energy for the cluster under the assumption that $R_{\mathrm{h},\mathrm{cl}}/\dot{R}_{\mathrm{h},\mathrm{cl}} = t_\mathrm{cl}$, $\mathrm{D}[\Delta E_\mathrm{cl}]\Big|_\mathrm{M}$ and $\mathrm{D}[\Delta E_\mathrm{cl}]\Big|_\mathrm{C}$ are the diffusion coefficients due to MACHOs and cold, collisionless dark matter, respectively, $\alpha$ and $\beta$ are constants depending on the density profile of the cluster, $\rho$ is the dark-matter density at the location of the cluster, and $t_\mathrm{cl}$ is the current age of the cluster.
    The diffusion coefficients are given in the appendix and depend on a number of variables, most importantly the three-dimensional velocity dispersions $\sigma_*$ of the stars and $\sigma_\mathrm{DM}$ of the dark matter.
    Any combination of parameters that leads to $f_\mathrm{M} > f_{\mathrm{M},\mathrm{lim}}$ can be rejected.

    For the star cluster's age~$t_\mathrm{cl}$ we adopt $8\,\mathrm{Gyr}$, the midpoint in the range of fitting isochrones.
    We also consider the lower extreme, $7\,\mathrm{Gyr}$, as an alternative, as this will give weaker constraints.
    For each age we use the corresponding IMF-based mass estimate of $M_\mathrm{cl}$.
    The typical stellar (remnant) mass is assumed to be $1\,M_\sun$, but the exact value is not important for the most interesting regime where the MACHO mass is much larger than stellar masses.
    The cluster's projected half-light radius~$R_{\mathrm{h},\mathrm{cl}}$ was determined by \citet{Crnojevic-2016-ApJL-824-L14} to be $13\,\mathrm{pc}$ and we adopt $\alpha = 0.37$ and $\beta = 7.2$, as calculated --~though not used~-- by \citet{Brandt-2016-ApJL-824-L31} based on the photometry by \citet{Crnojevic-2016-ApJL-824-L14}.
    The cluster's mass is estimated from the absolute V-band magnitude calculated by \citet{Crnojevic-2016-ApJL-824-L14} and the mass-to-light ratio of ${\sim}1.56$ that we determined in Sect.~\ref{ssec:properties}.
    Assuming a constant density of dark matter and stars in the central area of Eri~2, dominated by dark matter, one can derive from \citet[Eq.~9]{Wolf-2010-MNRAS-406-1220} that the mass inside a three-dimensional radius~$r$ is given by
    \begin{equation}
        M(r) = \frac{\sigma_\mathrm{DM}^2r}{G}.
    \end{equation}
    This leads to a density of
    \begin{equation}
        \rho(r) = \frac{3\sigma_\mathrm{DM}^2}{4\pi Gr^2}.
    \end{equation}
    We therefore estimate the dark-matter density in the centre of Eri~2 with the above equation, evaluated at the approximate radius of our field of view, $r = 50\,\mathrm{pc}$.
    The host galaxy Eri~2 is dark-matter dominated, so assuming dynamical equilibrium its stellar kinematics approximately trace the dark-matter distribution.
    We therefore have $\sigma_* = \sqrt{3}\sigma_\mathrm{cl}$ and $\sigma_\mathrm{DM} = \sqrt{3}\sigma_\mathrm{gal}$, taking into account the difference between line-of-sight and three-dimensional velocity dispersions.
    This gives an enclosed mass of $3.7^{+3.3}_{-1.9} \times 10^6\,M_\sun$ and a dark-matter density $\rho = 7.1^{+6.4}_{-3.7}\,M_\sun\,\mathrm{pc}^{-3}$, which is much higher than the value $0.15\,M_\sun\,\mathrm{pc}^{-3}$ found by \citet{Li-2017-ApJ-838-8}.
    This is because we find a similar velocity dispersion for the bulk of Eri~2, but at a much smaller radius.
    Now we are able to calculate the limiting MACHO fraction for a given MACHO mass and observed velocity dispersions.

\subsection{Constraints on massive compact halo objects as dark matter}
\label{ssec:constraints}
\begin{figure*}
        \includegraphics[width=1.0\linewidth]{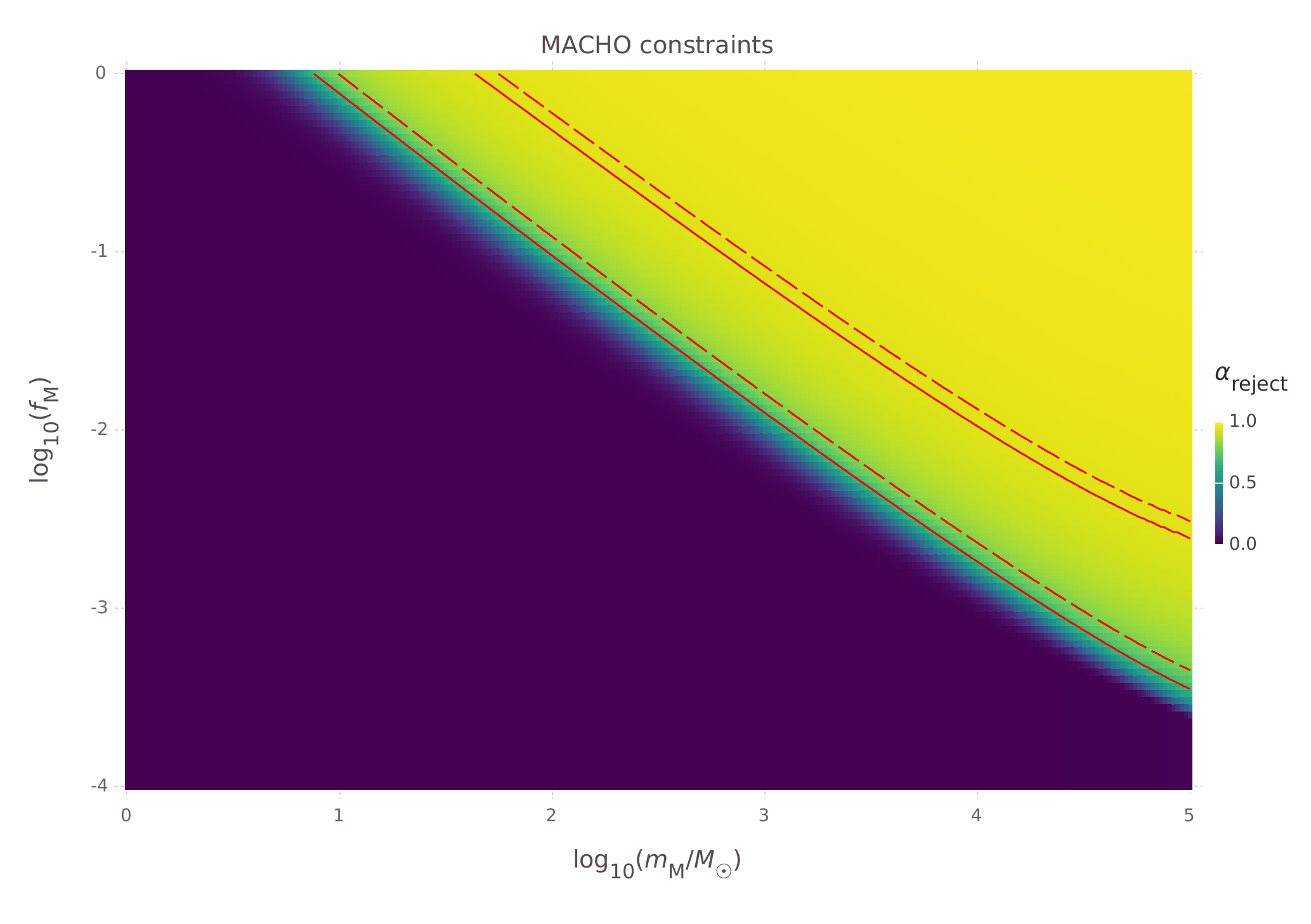}%
        \caption{%
            Constraints on massive astrophysical compact halo objects~(MACHOs) as a form of dark matter, based on the velocity distributions in Eridanus~2.
            For each combination of MACHO mass~$m_\mathrm{M}$ and the fraction~$f_\mathrm{M}$ of dark matter consisting of MACHOs, the probability of rejection by the observations is indicated, using the fiducial cluster age of $8\,\mathrm{Gyr}$.
            The red solid contours show where this probability reaches a 68- and 95-$\%$ level.
            For dark matter consisting entirely of MACHOs, $m_\mathrm{M} > 10^{0.88}\,M_\sun \approx 7.6\,M_\sun$ and $m_\mathrm{M} > 10^{1.64}\,M_\sun \approx 44\,M_\sun$ are ruled out at the 68- and 95-$\%$ level.
            The red dashed contours show how the constraints weaken slightly when a lower cluster age of $7\,\mathrm{Gyr}$ is assumed.
        }
        \label{fig:machos}
    \end{figure*}

    In reality, we do not know the exact value of the velocity dispersions.
    We will therefore have to integrate over the posterior likelihood distributions for the dispersions.
    We calculate the limiting MACHO fraction on a three-dimensional grid with the same velocity dispersion grid points as in the determination of the velocity distributions, as well as $251$ logarithmically spaced MACHO masses in the range $1$--$10^5\,M_\sun$.
    We then compare the results with a grid of $101$ logarithmically spaced MACHO fractions in the range $10^{-4}$--$1$.
    On the resulting four-dimensional grid of $\sigma_\mathrm{gal}$, $\sigma_\mathrm{cl}$, $m_\mathrm{M}$, and $f_\mathrm{M}$, we reject grid points if the MACHO fraction is higher than the calculated limiting MACHO fraction.
    We can calculate the `rejection level' for a given combination of MACHO fraction and mass by summing over the rejected grid points along the velocity dispersion axes, weighted by the likelihoods of the associated velocity dispersions:
    \begin{equation}
        \alpha_\mathrm{reject}(f_\mathrm{M}, m_\mathrm{M}) = \frac{\sum_{\sigma_\mathrm{gal},\sigma_\mathrm{cl}}^{f_\mathrm{M}>f_{\mathrm{M},\mathrm{lim}}(m_\mathrm{M}, \sigma_\mathrm{gal}, \sigma_\mathrm{cl})} \mathcal{L}(\sigma_\mathrm{gal}) \cdot \mathcal{L}(\sigma_\mathrm{cl})}{\sum_{\sigma_\mathrm{gal},\sigma_\mathrm{cl}} \mathcal{L}(\sigma_\mathrm{gal}) \cdot \mathcal{L}(\sigma_\mathrm{cl})}.
    \end{equation}
    The rejection level as a function of MACHO mass and fraction is show in Fig.~\ref{fig:machos}.
    At a MACHO fraction of unity, we can rule out $m_\mathrm{M} > 10^{0.88}\,M_\sun \approx 7.6\,M_\sun$ or $m_\mathrm{M} > 10^{1.64}\,M_\sun \approx 44\,M_\sun$ at the 68- or 95-$\%$ level, respectively.
    This means that, under the assumptions that we have made, dark matter cannot purely consist of MACHOs above these masses, at the stated confidence levels.
    For dark matter that only partly consists of MACHOs, the constraints become considerably weaker.
    Constraints for several values of~$f_\mathrm{M}$ are listed in Table~\ref{tab:machos}.
    \begin{table}
        \caption{%
            Rejected MACHO masses~$m_\mathrm{M}$ for different initial mass functions~(IMFs) and cluster ages at different MACHO fractions~$f_\mathrm{M}$, based on the velocity distributions in Eridanus~2.
            The standard Salpeter and Kroupa IMFs, as well as a Kroupa IMF modified to match the observed IMF in two other ultra-faint dwarf galaxies, provide the same (fiducial) results.
            Modifying the Salpeter IMF in the same way leads to slightly weaker constraints.
            The fiducial cluster age is $8\,\mathrm{Gyr}$; the lowest estimate of the cluster age, $7\,\mathrm{Gyr}$, gives slightly weaker constraints.
            The constraints are given for two values of $\alpha_\mathrm{reject}$, the confidence level for the model to be rejected by the data.
        }
        \label{tab:machos}
        \centering
        \begin{tabular}{ccccc}
            \hline\hline
            \hspace{7mm} & $\log_{10}(f_\mathrm{M})$ & \multicolumn{2}{c}{$\log_{10}(m_\mathrm{M}/M_\sun)$} & \hspace{7mm} \\
            & & ($\alpha_\mathrm{reject} > 0.68$) & ($\alpha_\mathrm{reject} > 0.95$) & \\
            \hline
            \multicolumn{5}{c}{Stand. Salpeter / stand. Kroupa / modif. Kroupa, fiducial age} \\
            \hline
            & $0$ & ${>}0.88$ & ${>}1.64$ & \\
            & $-1$ & ${>}1.98$ & ${>}2.80$ & \\
            & $-2$ & ${>}3.12$ & ${>}4.04$ & \\
            & $-3$ & ${>}4.34$ & --- & \\
            \hline
            \multicolumn{5}{c}{Modif. Salpeter, fiducial age} \\
            \hline
            & $0$ & ${>}0.90$ & ${>}1.68$ & \\
            & $-1$ & ${>}2.00$ & ${>}2.82$--$2.84$\tablefootmark{a} & \\
            & $-2$ & ${>}3.14$ & ${>}4.08$ & \\
            & $-3$ & ${>}4.36$ & --- & \\
            \hline
            \multicolumn{5}{c}{Stand. Salpeter / stand. Kroupa / modif. Kroupa, lowest age} \\
            \hline
            & $0$ & ${>}1.00$ & ${>}1.76$ & \\
            & $-1$ & ${>}2.10$ & ${>}2.92$ & \\
            & $-2$ & ${>}3.24$ & ${>}4.16$--$4.18$\tablefootmark{a} & \\
            & $-3$ & ${>}4.48$ & --- & \\
            \hline
            \multicolumn{5}{c}{Modif. Salpeter, lowest age} \\
            \hline
            & $0$ & ${>}1.02$ & ${>}1.78$--$1.80$\tablefootmark{a} & \\
            & $-1$ & ${>}2.12$ & ${>}2.94$ & \\
            & $-2$ & ${>}3.26$ & ${>}4.20$--$4.22$\tablefootmark{a} & \\
            & $-3$ & ${>}4.50$ & --- & \\
            \hline
        \end{tabular}
        \tablefoot{%
            \tablefoottext{a}{Value varies slightly over metallicity range of $[\mathrm{Fe/H}] = -3.8$ to $-1.0$.}
        }
    \end{table}

\section{Discussion}
\label{sec:discussion}
    Our 68-$\%$ contour is comparable to the result derived by \citet{Li-2017-ApJ-838-8}, even though we find a much larger dark-matter density.
    This is because $f_{\mathrm{M},\mathrm{lim}}$ mainly depends on $\sigma_\mathrm{DM}$, which is similar to previous results, rather than on $\rho$.
    The 95-$\%$ constraint is much weaker, but this is to be expected as we allow for 128 different scenarios of cluster membership and have propagated this through the velocity dispersions into the MACHO constraints.
    Other constraints of various strengths do exist in this mass range, from lack of mass segregation in Segue~1~\citep{Koushiappas-2017-PhRvL-119-041102}, the distribution of wide-binary separations~\citep{MonroyRodriguez-2014-ApJ-790-159}, and lack of an imprint on the cosmic microwave background from accreting primordial black holes~\citep{AliHaimoud-2017-PhRvD-95-043534}.
    Another constraint~\citep{Stegmann-arXiv-1910}, comparing the observed half-light radii of a large sample of UFDs to simulations, also rules out this mass range for dark matter consisting completely of primordial black holes, even when considering extended mass distributions, but the methodology is unable to provide answers for lower abundances.
    As each of these astrophysical constraints involves assumptions, corroboration from different sources is still valuable, especially considering this mass range is the last window for dark matter purely consisting of MACHOs that is not robustly closed by more fundamental physics.

    Several estimates and assumptions enter into the calculation of the MACHO constraints.
    The cluster's mass is difficult to establish from the kinematics, which is why we have estimated it from photometry.
    This, however, depends on the mass-to-light ratio, which depends on the choice of initial mass function and the choice of isochrone.
    For the standard Salpeter and Kroupa IMFs and the modified Kroupa IMF described in Sect.~\ref{ssec:properties}, for any value of metallicity between $-3.8\,\mathrm{dex}$ and $-1.0\,\mathrm{dex}$ we find the same rejection limits at $f_\mathrm{M} = 1$.
    The modified Salpeter IMF, with a tenfold larger mass-to-light ratio, only slightly increases these limits.
    Table~\ref{tab:machos} lists the results for all the different choices of IMF we have tried.
    Additionally, one could argue about the most appropriate mass estimator for the dark-matter mass in the central region of Eri~2.
    All common mass estimators differ only in a scalar pre-factor.
    We have tested the influence of this pre-factor by running our analysis with a dark-matter mass differing by a factor two from the fiducial value.
    This had an influence of at most $0.02\,\mathrm{dex}$ on the limiting MACHO masses, which is the resolution of our parameter grid.

    We have assumed the cluster has the same age as determined for the bulk of Eri~2: $8\,\mathrm{Gyr}$.
    The photometry of cluster stars does seem to be consistent with the isochrones for the bulk, but we cannot be more precise because we have too few cluster stars to fit an isochrone and it is therefore outside the scope of this paper.
    Our lower limit of the age of Eri~2, $7\,\mathrm{Gyr}$, results in a slightly higher value of $f_{\mathrm{M},\mathrm{lim}}$ (see Table~\ref{tab:machos}).
    This results in a more conservative constraint, meaning a smaller part of the parameter space is rejected in this case.

    With only 26~stars, the 68-$\%$ confidence limit is affected by sampling variance.
    We have observed that, when changing the spexxy velocity measurements for those of ULySS or and older version of spexxy, it can shift by a factor of a few.
    On the other hand, the 95-$\%$ MACHO constraints have changed by at most ${\sim}10\,\%$ and are therefore robust against sampling variance.

    The insensitivity of our result to the mass-to-light ratio and dynamical-mass estimator can be understood by examining the behaviour of $f_{\mathrm{M},\mathrm{lim}}$ with respect to $M_\mathrm{cl}$ and $\rho$ in leading order:
    \begin{equation}
        f_{\mathrm{M},\mathrm{lim}} \sim \frac{\dot{U}_{\mathrm{cl},\mathrm{lim}}}{\rho M_\mathrm{cl}} \sim \frac{\alpha M_\mathrm{cl}}{\rho R_{\mathrm{h},\mathrm{cl}}} + 2\beta R_{\mathrm{h},\mathrm{cl}}^2.
    \end{equation}
    The ratio of these terms is on the order of the ratio of the cluster's stellar mass to the dark-matter mass enclosed in the half-light radius of the cluster.
    The second term clearly dominates, therefore it is understandable that changes to $M_\mathrm{cl}$ through the mass-to-light ratio or $\rho$ through the mass estimator do not have a large effect on the MACHO constraints.

    Similarly, when studying the behaviour of $f_{\mathrm{M},\mathrm{lim}}$ with respect to the velocity dispersions, we can see that it scales linearly with them to first order.
    Therefore, if there are unresolved binary stars in our sample, which would inflate our estimates of the velocity dispersions, we have overestimated the limiting MACHO abundance.
    Our exclusion of the parameter space is therefore also conservative in this respect.

    For the estimation of the dark-matter density from the stellar velocity dispersion, we had to assume Eridanus~2 is in dynamical equilibrium.
    Considering this dwarf galaxy is at a large distance from the Milky Way and that it is currently near its pericentre~\citep{Fritz-2018-A&A-619-A103}, the tidal influence from the Milky Way is small.
    Given its age, we argue the system has had sufficient time to reach dynamical equilibrium.

    In deriving the MACHO constraints, we have used a diffusion approximation.
    This may not be valid for large MACHO masses or very small numbers of MACHOs.
    A MACHO with a mass comparable to the star cluster might, instead of dynamically heat, tidally strip or deform the cluster, giving a much larger effect than calculated here.
    On the other hand, if the number of MACHOs is too small, which is also a consequence of large MACHO masses, the cluster might never encounter a MACHO within its lifetime.
    In this case, our constraints may be too strong.
    Because these two considerations are opposite in effect, it is difficult to pin down a limit to the diffusion approximation.
    The majority of interactions will be long-distance; as the mass inside $50\,\mathrm{pc}$ of the cluster is ${\sim}10^6\,M_\sun$, we advise caution when working with MACHO masses ${\gtrsim}10^4\,M_\sun$.
    In that regime other constraints from wide binaries~\citep{MonroyRodriguez-2014-ApJ-790-159} and the cosmic microwave background~\citep{AliHaimoud-2017-PhRvD-95-043534} exist.

    Though we only consider a MACHO population with a single mass, it has been shown~\citep{Carr-2017-PhRvD-96-023514} that MACHO constraints can become more stringent if one considers mass distributions.
    To properly study which mass distributions are allowed, single-mass MACHO constraints are required over the entire possible MACHO mass range.
    This is beyond the scope of this article, so we leave it to others to use our single-mass constraint in a meta-analysis.

\section{Conclusions}
\label{sec:conclusions}
    We find spectroscopic evidence in favour of the existence of a stellar cluster in the ultra-faint dwarf galaxy Eridanus~2.
    We determine intrinsic mean line-of-sight velocities $79.7^{+3.1}_{-3.8}\,\mathrm{km}\,\mathrm{s}^{-1}$ and $76.0^{+3.2}_{-3.7}\,\mathrm{km}\,\mathrm{s}^{-1}$ for the putative stellar cluster and the bulk of Eridanus~2, respectively, and intrinsic velocity dispersions ${<}7.6\,\mathrm{km}\,\mathrm{s}^{-1}$ (68-$\%$ upper limit) or ${<}17.5\,\mathrm{km}\,\mathrm{s}^{-1}$ (95-$\%$ upper limit) for the cluster and $10.3^{+3.9}_{-3.2}\,\mathrm{km}\,\mathrm{s}^{-1}$ for the bulk, based on observations from MUSE-Faint, a survey of ultra-faint dwarf galaxies with the Multi Unit Spectroscopic Explorer on the Very Large Telescope, of 26~member stars in the central square arcminute of Eridanus~2.
    These velocity distributions are consistent with earlier research and, combined with existing photometry, show that the existence of the cluster is likely.
    This would make Eridanus~2 the lowest-mass galaxy hosting a stellar cluster.
    We also find evidence for the presence of carbon stars in Eridanus~2.
    The existence of such stars would indicate there is, aside from the old population, a second, intermediate-age population.

    Assuming the existence of the cluster, we derive constraints on what fraction of dark matter could be massive astrophysical compact halo objects with masses between $1$--$10^5\,M_\sun$.
    We find that, for dark matter completely made of massive astrophysical compact halo objects, masses larger than $10^{0.88}\,M_\sun \approx 7.6\,M_\sun$ and $10^{1.64}\,M_\sun \approx 44\,M_\sun$ are rejected at a 68- and 95-$\%$ level, respectively.
    These constraints are less strict at lower MACHO abundances.
    This result is robust against the choice of initial mass function, variations in metallicity, and the choice of dynamical-mass estimator.
    The diffusion approximation may also influence the constraints at the extremes of the parameter space, particularly above a MACHO mass of ${\sim}10^4\,M_\sun$.

    \begin{acknowledgements}
        We wish to thank the referee, Timothy~D. Brandt, for his constructive report and helpful comments.
        We wish to thank Simon Conseil for providing us with the bad-pixel mask from the MUSE \emph{Hubble} Ultra Deep Field Survey and Peter Weilbacher for providing early access to version~2.4 of the MUSE Data Reduction Software with the newly integrated autocalibration method.
        SLZ wishes to thank David~J.E. Marsh and Jens~C. Niemeyer for helpful discussions and feedback.

        We wish to acknowledge the use of several open-source software libraries and collections that have been particularly helpful in the production of this paper, namely Astropy~\citep{Robitaille-2013-A&A-558-A33}, corner.py~\citep{ForemanMackey-2016-JOSS-1-24}, emcee~\citep{ForemanMackey-2013-PASP-125-306}, Gadfly\footnote{\url{http://gadflyjl.org/}}, Julia Astro\footnote{\url{http://juliaastro.github.io/}}, and the SciPy ecosystem\footnote{\url{https://scipy.org/}}.

        SLZ acknowledges support by The Netherlands Organisation for Scientific Research~(NWO) through a TOP Grant Module~1 under project number 614.001.652.
        JB acknowledges support by FCT/MCTES through national funds by the grant UID/FIS/04434/2019 and through the Investigador FCT Contract No.~IF/01654/2014/CP1215/CT0003.
        MLPG acknowledges the support from the European Union's Horizon 2020 research and innovation programme under the Marie Sklodowska-Curie grant agreement No~707693.
        TOH, MMR, MdB, SD, and DK acknowledge support from the German Ministry for Education and Science (BMBF Verbundforschung) through grants 05A14MGA, 05A17MGA, 05A14BAC, and 05A17BAA.
        SK gratefully acknowledges funding from a European Research Council consolidator grant (ERC-CoG-646928-Multi-Pop).
        This research is supported by the German Research Foundation~(DFG) with grants DR~281/35-1 and KA~4537/2-1.
        Based on observations made with ESO Telescopes at the La Silla Paranal Observatory under programme ID 0100.D-0807.
    \end{acknowledgements}

    \bibliographystyle{aa}
    \bibliography{Zoutendijk_Eridanus2-cluster}

\begin{thebibliography}{76}
\expandafter\ifx\csname natexlab\endcsname\relax\def\natexlab#1{#1}\fi

\bibitem[{Abbott {et~al.}(2016)Abbott, Abbott, Abbott, Abernathy, Acernese,
  Ackley, Adams, Adams, Addesso, Adhikari, Adya, Affeldt, Agathos, Agatsuma,
  Aggarwal, Aguiar, Aiello, Ain, Ajith, Allen, Allocca, Altin, Anderson,
  Anderson, Arai, Arain, Araya, Arceneaux, Areeda, Arnaud, Arun, Ascenzi,
  Ashton, Ast, Aston, Astone, Aufmuth, Aulbert, Babak, Bacon, Bader, Baker,
  Baldaccini, Ballardin, Ballmer, Barayoga, Barcay, Barish, Barker, Barone,
  Barr, Barsotti, Barsuglia, Barta, Bartlett, Barton, Bartos, Bassiri, Basti,
  Batch, Baune, Bavigadda, Bazzan, Behnke, Bejger, Belczynski, Bell, Bell,
  Berger, Bergman, Bergmann, Berry, Bersanetti, Bertolini, Betzwieser, Bhagwat,
  Bhandare, Bilenko, Billingsley, Birch, Birney, Birnholtz, Biscans, Bisht,
  Bitossi, Biwer, Bizouard, Blackburn, Blair, Blair, Blair, Bloemen, Bock,
  Bodiya, Boer, Bogaert, Bogan, Bohe, Bojtos, Bond, Bondu, Bonnand, Boom, Bork,
  Boschi, Bose, Bouffanais, Bozzi, Bradaschia, Brady, Braginsky, Branchesi,
  Brau, Briant, Brillet, Brinkmann, Brisson, Brockill, Brooks, Brown, Brown,
  Brown, Buchanan, Buikema, Bulik, Bulten, Buonanno, Buskulic, Buy, Byer,
  Cabero, Cadonati, Cagnoli, Cahillane, Bustillo, Callister, Calloni, Camp,
  Cannon, Cao, Capano, Capocasa, Carbognani, Caride, Diaz, Casentini, Caudill,
  Cavaglia, Cavalier, Cavalieri, Cella, Cepeda, Baiardi, Cerretani, Cesarini,
  Chakraborty, Chalermsongsak, Chamberlin, Chan, Chao, Charlton,
  Chassande-Mottin, Chen, Chen, Cheng, Chincarini, Chiummo, Cho, Cho, Chow,
  Christensen, Chu, Chua, Chung, Ciani, Clara, Clark, Cleva, Coccia, Cohadon,
  Colla, Collette, Cominsky, Constancio, Conte, Conti, Cook, Corbitt, Cornish,
  Corsi, Cortese, Costa, Coughlin, Coughlin, Coulon, Countryman, Couvares,
  Cowan, Coward, Cowart, Coyne, Coyne, Craig, Creighton, Creighton, Cripe,
  Crowder, Cruise, Cumming, Cunningham, Cuoco, Canton, Danilishin, D'Antonio,
  Danzmann, Darman, Costa, Dattilo, Dave, Daveloza, Davier, Davies, Daw, Day,
  De, DeBra, Debreczeni, Degallaix, Laurentis, Deleglise, Pozzo, Denker, Dent,
  Dereli, Dergachev, DeRosa, Rosa, DeSalvo, Dhurandhar, Diaz, Fiore, Giovanni,
  Lieto, Pace, Palma, Virgilio, Dojcinoski, Dolique, Donovan, Dooley, Doravari,
  Douglas, Downes, Drago, Drever, Driggers, Du, Ducrot, Dwyer, Edo, Edwards,
  Effler, Eggenstein, Ehrens, Eichholz, Eikenberry, Engels, Essick, Etzel,
  Evans, Evans, Everett, Factourovich, Fafone, Fair, Fairhurst, Fan, Fang,
  Farinon, Farr, Farr, Favata, Fays, Fehrmann, Fejer, Feldbaum, Ferrante,
  Ferreira, Ferrini, Fidecaro, Finn, Fiori, Fiorucci, Fisher, Flaminio,
  Fletcher, Fong, Fournier, Franco, Frasca, Frasconi, Frede, Frei, Freise,
  Frey, Frey, Fricke, Fritschel, Frolov, Fulda, Fyffe, Gabbard, Gair,
  Gammaitoni, Gaonkar, Garufi, Gatto, Gaur, Gehrels, Gemme, Gendre, Genin,
  Gennai, George, Gergely, Germain, Ghosh, Ghosh, Ghosh, Giaime, Giardina,
  Giazotto, Gill, Glaefke, Gleason, Goetz, Goetz, Gondan, Gonzalez, Castro,
  Gopakumar, Gordon, Gorodetsky, Gossan, Gosselin, Gouaty, Graef, Graff,
  Granata, Grant, Gras, Gray, Greco, Green, Greenhalgh, Groot, Grote,
  Grunewald, Guidi, Guo, Gupta, Gupta, Gushwa, Gustafson, Gustafson, Hacker,
  Hall, Hall, Hammond, Haney, Hanke, Hanks, Hanna, Hannam, Hanson, Hardwick,
  Harms, Harry, Harry, Hart, Hartman, Haster, Haughian, Healy, Heefner,
  Heidmann, Heintze, Heinzel, Heitmann, Hello, Hemming, Hendry, Heng, Hennig,
  Heptonstall, Heurs, Hild, Hoak, Hodge, Hofman, Hollitt, Holt, Holz, Hopkins,
  Hosken, Hough, Houston, Howell, Hu, Huang, Huerta, Huet, Hughey, Husa,
  Huttner, Huynh-Dinh, Idrisy, Indik, Ingram, Inta, Isa, Isac, Isi, Islas,
  Isogai, Iyer, Izumi, Jacobson, Jacqmin, Jang, Jani, Jaranowski, Jawahar,
  Jimenez-Forteza, Johnson, Johnson-McDaniel, Jones, Jones, Jonker, Ju, Haris,
  Kalaghatgi, Kalogera, Kandhasamy, Kang, Kanner, Karki, Kasprzack,
  Katsavounidis, Katzman, Kaufer, Kaur, Kawabe, Kawazoe, Kefelian, Kehl,
  Keitel, Kelley, Kells, Kennedy, Keppel, Key, Khalaidovski, Khalili, Khan,
  Khan, Khan, Khazanov, Kijbunchoo, Kim, Kim, Kim, Kim, Kim, Kim, King, King,
  Kinzel, Kissel, Kleybolte, Klimenko, Koehlenbeck, Kokeyama, Koley,
  Kondrashov, Kontos, Koranda, Korobko, Korth, Kowalska, Kozak, Kringel,
  Krishnan, Krolak, Krueger, Kuehn, Kumar, Kumar, Kuo, Kutynia, Kwee, Lackey,
  Landry, Lange, Lantz, Lasky, Lazzarini, Lazzaro, Leaci, Leavey, Lebigot, Lee,
  Lee, Lee, Lee, Lenon, Leonardi, Leong, Leroy, Letendre, Levin, Levine, Li,
  Libson, Littenberg, Lockerbie, Logue, Lombardi, London, Lord, Lorenzini,
  Loriette, Lormand, Losurdo, Lough, Lousto, Lovelace, Luck, Lundgren, Luo,
  Lynch, Ma, MacDonald, Machenschalk, MacInnis, Macleod, Magana-Sandoval,
  Magee, Mageswaran, Majorana, Maksimovic, Malvezzi, Man, Mandel, Mandic,
  Mangano, Mansell, Manske, Mantovani, Marchesoni, Marion, Marka, Marka,
  Markosyan, Maros, Martelli, Martellini, Martin, Martin, Martynov, Marx,
  Mason, Masserot, Massinger, Masso-Reid, Matichard, Matone, Mavalvala,
  Mazumder, Mazzolo, McCarthy, McClelland, McCormick, McGuire, McIntyre,
  McIver, McManus, McWilliams, Meacher, Meadors, Meidam, Melatos, Mendell,
  Mendoza-Gandara, Mercer, Merilh, Merzougui, Meshkov, Messenger, Messick,
  Meyers, Mezzani, Miao, Michel, Middleton, Mikhailov, Milano, Miller,
  Millhouse, Minenkov, Ming, Mirshekari, Mishra, Mitra, Mitrofanov,
  Mitselmakher, Mittleman, Moggi, Mohan, Mohapatra, Montani, Moore, Moore,
  Moraru, Moreno, Morriss, Mossavi, Mours, Mow-Lowry, Mueller, Mueller, Muir,
  Mukherjee, Mukherjee, Mukherjee, Mukund, Mullavey, Munch, Murphy, Murray,
  Mytidis, Nardecchia, Naticchioni, Nayak, Necula, Nedkova, Nelemans, Neri,
  Neunzert, Newton, Nguyen, Nielsen, Nissanke, Nitz, Nocera, Nolting,
  Normandin, Nuttall, Oberling, Ochsner, O'Dell, Oelker, Ogin, Oh, Oh, Ohme,
  Oliver, Oppermann, Oram, O'Reilly, O'Shaughnessy, Ott, Ottaway, Ottens,
  Overmier, Owen, Pai, Pai, Palamos, Palashov, Palomba, Pal-Singh, Pan, Pan,
  Pankow, Pannarale, Pant, Paoletti, Paoli, Papa, Paris, Parker, Pascucci,
  Pasqualetti, Passaquieti, Passuello, Patricelli, Patrick, Pearlstone,
  Pedraza, Pedurand, Pekowsky, Pele, Penn, Perreca, Pfeiffer, Phelps, Piccinni,
  Pichot, Pickenpack, Piergiovanni, Pierro, Pillant, Pinard, Pinto, Pitkin,
  Poeld, Poggiani, Popolizio, Post, Powell, Prasad, Predoi, Premachandra,
  Prestegard, Price, Prijatelj, Principe, Privitera, Prix, Prodi, Prokhorov,
  Puncken, Punturo, Puppo, Purrer, Qi, Qin, Quetschke, Quintero, Quitzow-James,
  Raab, Rabeling, Radkins, Raffai, Raja, Rakhmanov, Ramet, Rapagnani, Raymond,
  Razzano, Re, Read, Reed, Regimbau, Rei, Reid, Reitze, Rew, Reyes, Ricci,
  Riles, Robertson, Robie, Robinet, Rocchi, Rolland, Rollins, Roma, Romano,
  Romano, Romanov, Romie, Rosinska, Rowan, Rudiger, Ruggi, Ryan, Sachdev,
  Sadecki, Sadeghian, Salconi, Saleem, Salemi, Samajdar, Sammut, Sampson,
  Sanchez, Sandberg, Sandeen, Sanders, Sanders, Sassolas, Sathyaprakash,
  Saulson, Sauter, Savage, Sawadsky, Schale, Schilling, Schmidt, Schmidt,
  Schnabel, Schofield, Schonbeck, Schreiber, Schuette, Schutz, Scott, Scott,
  Sellers, Sengupta, Sentenac, Sequino, Sergeev, Serna, Setyawati, Sevigny,
  Shaddock, Shaffer, Shah, Shahriar, Shaltev, Shao, Shapiro, Shawhan, Sheperd,
  Shoemaker, Shoemaker, Siellez, Siemens, Sigg, Silva, Simakov, Singer, Singer,
  Singh, Singh, Singhal, Sintes, Slagmolen, Smith, Smith, Smith, Smith, Son,
  Sorazu, Sorrentino, Souradeep, Srivastava, Staley, Steinke, Steinlechner,
  Steinlechner, Steinmeyer, Stephens, Stevenson, Stone, Strain, Straniero,
  Stratta, Strauss, Strigin, Sturani, Stuver, Summerscales, Sun, Sutton,
  Swinkels, Szczepanczyk, Tacca, Talukder, Tanner, Tapai, Tarabrin, Taracchini,
  Taylor, Theeg, Thirugnanasambandam, Thomas, Thomas, Thomas, Thorne, Thorne,
  Thrane, Tiwari, Tiwari, Tokmakov, Tomlinson, Tonelli, Torres, Torrie, Toyra,
  Travasso, Traylor, Trifiro, Tringali, Trozzo, Tse, Turconi, Tuyenbayev,
  Ugolini, Unnikrishnan, Urban, Usman, Vahlbruch, Vajente, Valdes, Vallisneri,
  van Bakel, van Beuzekom, van~den Brand, Broeck, Vander-Hyde, van~der Schaaf,
  van Heijningen, van Veggel, Vardaro, Vass, Vasuth, Vaulin, Vecchio, Vedovato,
  Veitch, Veitch, Venkateswara, Verkindt, Vetrano, Vicere, Vinciguerra, Vine,
  Vinet, Vitale, Vo, Vocca, Vorvick, Voss, Vousden, Vyatchanin, Wade, Wade,
  Wade, Waldman, Walker, Wallace, Walsh, Wang, Wang, Wang, Wang, Wang, Ward,
  Ward, Warner, Was, Weaver, Wei, Weinert, Weinstein, Weiss, Welborn, Wen,
  Wessels, Westphal, Wette, Whelan, Whitcomb, White, Whiting, Wiesner,
  Wilkinson, Willems, Williams, Williams, Williamson, Willis, Willke, Wimmer,
  Winkelmann, Winkler, Wipf, Wiseman, Wittel, Woan, Worden, Wright, Wu, Yablon,
  Yakushin, Yam, Yamamoto, Yancey, Yap, Yu, Yvert, Zadrozny, Zangrando,
  Zanolin, Zendri, Zevin, Zhang, Zhang, Zhang, Zhang, Zhao, Zhou, Zhou, Zhu,
  Zucker, Zuraw, \& Zweizig}]{Abbott-2016-PhRvL-116-061102}
Abbott, B.~P., Abbott, R., Abbott, T.~D., {et~al.} 2016, \prl, 116, 061102

\bibitem[{Abbott {et~al.}(2005)Abbott, Aldering, Annis, Barlow, Bebek, Bigelow,
  Beldica, Bernstein, Bridle, Brunner, Carlstrom, Campbell, Castander, Cunha,
  Diehl, Dodelson, Doel, Efstathiou, Estrada, Evrard, Fern\'andez, Flaugher,
  Fosalba, Frieman, Gazta{\~n}aga, Gerdes, Gladders, Hu, Huterer, Jain,
  Karliner, Kent, Lahav, Levi, Lima, Lin, Limon, Mart\'{\i}nez, McKay, McMahon,
  Merritt, Miller, Miralda-Escude, Mohr, Nichol, Oyaizu, Peacock, Peoples,
  Perlmutter, Plante, Ricker, Roe, Scarpine, Schubnell, Selen, Sheldon, Smith,
  Stebbins, Stoughton, andW. Sutherland, Takada, Tarle, Tecchio, Thaler,
  Tucker, Viti, Walker, Wechsler, Weller, \& Wester}]{Abbott-arXiv-0510}
Abbott, T., Aldering, G., Annis, J., {et~al.} 2005, arXiv e-prints
  [\eprint[arXiv]{astro-ph/0510346}]

\bibitem[{Ali-Ha\"{\i}moud \&
  Kamionkowski(2017)}]{AliHaimoud-2017-PhRvD-95-043534}
Ali-Ha\"{\i}moud, Y. \& Kamionkowski, M. 2017, \prd, 95, 043534

\bibitem[{Bacon {et~al.}(2010)Bacon, Accardo, Adjali, Anwand, Bauer, Biswas,
  Blaizot, Boudon, Brau-Nogue, Brinchmann, Caillier, Capoani, Carollo, Contini,
  Couderc, Daguis\'e, Deiries, Delabre, Dreizler, Dubois, Dupieux, Dupuy,
  Emsellem, Fechner, Fleischmann, Fran\c{c}ois, Gallou, Gharsa, Glindemann,
  Gojak, Guiderdoni, Hansali, Hahn, Jarno, Kelz, Koehler, Kosmalski, Laurent,
  {Le Floch}, Lilly, Lizon, Loupias, Manescau, Monstein, Nicklas, Olaya, Pares,
  Pasquini, P\'econtal-Rousset, Pell\'o, Petit, Popow, Reiss, Remillieux,
  Renault, Roth, Rupprecht, Serre, Schaye, Soucail, Steinmetz, Streicher,
  Stuik, Valentin, Vernet, Weilbacher, Wisotzki, \&
  Yerle}]{Bacon-2010-SPIE-7735-773508}
Bacon, R., Accardo, M., Adjali, L., {et~al.} 2010, \procspie, 7735, 773508

\bibitem[{Bacon {et~al.}(2017)Bacon, Conseil, Mary, Brinchmann, Shepherd,
  Akhlaghi, Weilbacher, Piqueras, Wisotzki, Lagattuta, Epinat, Guerou, Inami,
  Cantalupo, Courbot, Contini, Richard, Maseda, Bouwens, Bouche, Kollatschny,
  Schaye, Marino, Pello, Herenz, Guiderdoni, \&
  Carollo}]{Bacon-2017-A&A-608-A1}
Bacon, R., Conseil, S., Mary, D., {et~al.} 2017, \aap, 608, A1

\bibitem[{Bechtol {et~al.}(2015)Bechtol, Drlica-Wagner, Balbinot, Pieres,
  Simon, Yanny, Santiago, Wechsler, Frieman, Walker, Williams, Rozo, Rykoff,
  Queiroz, Luque, Benoit-L\'evy, Tucker, Sevilla, Gruendl, da~Costa, Neto,
  Maia, Abbott, Allam, Armstrong, Bauer, Bernstein, Bernstein, Bertin, Brooks,
  Buckley-Geer, Burke, Rosell, Castander, Covarrubias, D'Andrea, DePoy, Desai,
  Diehl, Eifler, Estrada, Evrard, Fernandez, Finley, Flaugher, Gaztanaga,
  Gerdes, Girardi, Gladders, Gruen, Gutierrez, Hao, Honscheid, Jain, James,
  Kent, Kron, Kuehn, Kuropatkin, Lahav, Li, Lin, Makler, March, Marshall,
  Martini, Merritt, Miller, Miquel, Mohr, Neilsen, Nichol, Nord, Ogando,
  Peoples, Petravick, Plazas, Romer, Roodman, Sako, Sanchez, Scarpine,
  Schubnell, Smith, Soares-Santos, Sobreira, Suchyta, Swanson, Tarle, Thaler,
  Thomas, Wester, \& Zuntz}]{Bechtol-2015-ApJ-807-50}
Bechtol, K., Drlica-Wagner, A., Balbinot, E., {et~al.} 2015, \apj, 807, 50

\bibitem[{Bertin \& Arnouts(1996)}]{Bertin-1996-A&AS-117-393}
Bertin, E. \& Arnouts, S. 1996, \aaps, 117, 393

\bibitem[{Bertone \& Tait(2018)}]{Bertone-2018-Natur-562-51}
Bertone, G. \& Tait, T. M.~P. 2018, \nat, 562, 51

\bibitem[{Binney \& Tremaine(2008)}]{Binney-2008-GD-PU-2}
Binney, J. \& Tremaine, S. 2008, Galactic Dynamics, 2nd edn. (Princeton, NJ:
  Princeton University)

\bibitem[{Bird {et~al.}(2016)Bird, Cholis, Mu{\~n}oz, Ali-Ha\"{\i}moud,
  Kamionkowski, Kovetz, Raccanelli, \& Riess}]{Bird-2016-PhRvL-116-201301}
Bird, S., Cholis, I., Mu{\~n}oz, J.~B., {et~al.} 2016, \prl, 116, 201301

\bibitem[{Brandt(2016)}]{Brandt-2016-ApJL-824-L31}
Brandt, T.~D. 2016, \apjl, 824, L31

\bibitem[{{Broadhurst} {et~al.}(2018){Broadhurst}, {Diego}, \&
  {Smoot}}]{Broadhurst-arXiv-1802}
{Broadhurst}, T., {Diego}, J.~M., \& {Smoot}, III, G. 2018, arXiv e-prints
  [\eprint[arXiv]{1802.05273}]

\bibitem[{Brooks \& Zolotov(2014)}]{Brooks-2014-ApJ-786-87}
Brooks, A.~M. \& Zolotov, A. 2014, \apj, 786, 87

\bibitem[{Bullock \& Boylan-Kolchin(2017)}]{Bullock-2017-ARA&A-55-343}
Bullock, J.~S. \& Boylan-Kolchin, M. 2017, \araa, 55, 343

\bibitem[{Carr {et~al.}(2017)Carr, Raidal, Tenkanen, Vaskonen, \&
  Veermae}]{Carr-2017-PhRvD-96-023514}
Carr, B., Raidal, M., Tenkanen, T., Vaskonen, V., \& Veermae, H. 2017, \prd,
  96, 023514

\bibitem[{Carr \& Hawking(1974)}]{Carr-1974-MNRAS-168-399}
Carr, B.~J. \& Hawking, S.~W. 1974, \mnras, 168, 399

\bibitem[{Choi {et~al.}(2016)Choi, Dotter, Conroy, Cantiello, Paxton, \&
  Johnson}]{Choi-2016-ApJ-823-102}
Choi, J., Dotter, A., Conroy, C., {et~al.} 2016, \apj, 823, 102

\bibitem[{Clesse \& Garc\'{\i}a-Bellido(2015)}]{Clesse-2015-PhRvD-92-023524}
Clesse, S. \& Garc\'{\i}a-Bellido, J. 2015, \prd, 92, 023524

\bibitem[{Contenta {et~al.}(2018)Contenta, Balbinot, Petts, Read, Gieles,
  Collins, Pe{\~n}arrubia, Delorme, \&
  Gualandris}]{Contenta-2018-MNRAS-476-3124}
Contenta, F., Balbinot, E., Petts, J.~A., {et~al.} 2018, \mnras, 476, 3124

\bibitem[{Crnojevi\'c {et~al.}(2016)Crnojevi\'c, Sand, Zaritsky, Spekkens,
  Willman, \& Hargis}]{Crnojevic-2016-ApJL-824-L14}
Crnojevi\'c, D., Sand, D.~J., Zaritsky, D., {et~al.} 2016, \apjl, 824, L14

\bibitem[{{Di Cintio} {et~al.}(2014){Di Cintio}, Brook, Macci{\`o}, Stinson,
  Knebe, Dutton, \& Wadsley}]{DiCintio-2014-MNRAS-437-415}
{Di Cintio}, A., Brook, C.~B., Macci{\`o}, A.~V., {et~al.} 2014, \mnras, 437,
  415

\bibitem[{Dotter(2016)}]{Dotter-2016-ApJS-222-8}
Dotter, A. 2016, \apjs, 222, 8

\bibitem[{Edelsbrunner {et~al.}(1983)Edelsbrunner, Kirkpatrick, \&
  Seidel}]{Edelsbrunner-1983-ITIT-29-551}
Edelsbrunner, H., Kirkpatrick, D.~G., \& Seidel, R. 1983, IEEE Trans. Inf.
  Theory, 29, 551

\bibitem[{Fitts {et~al.}(2019)Fitts, Boylan-Kolchin, Bozek, Bullock, Graus,
  Robles, Hopkins, El-Badry, Garrison-Kimmel, Faucher-Gigu{\`e}re, Wetzel, \&
  Kere{\v{s}}}]{Fitts-2019-MNRAS-490-962}
Fitts, A., Boylan-Kolchin, M., Bozek, B., {et~al.} 2019, \mnras, 490, 962

\bibitem[{Foreman-Mackey(2016)}]{ForemanMackey-2016-JOSS-1-24}
Foreman-Mackey, D. 2016, J. Open Source Softw., 1, 24

\bibitem[{Foreman-Mackey {et~al.}(2013)Foreman-Mackey, Hogg, Lang, \&
  Goodman}]{ForemanMackey-2013-PASP-125-306}
Foreman-Mackey, D., Hogg, D.~W., Lang, D., \& Goodman, J. 2013, \pasp, 125, 306

\bibitem[{Fritz {et~al.}(2018)Fritz, Bahaglia, Pawlowski, Kallivayalil, van~der
  Marel, Sohn, Brook, \& Besla}]{Fritz-2018-A&A-619-A103}
Fritz, T.~K., Bahaglia, G., Pawlowski, M.~S., {et~al.} 2018, \aap, 619, A103

\bibitem[{Geha {et~al.}(2013)Geha, Brown, Tumlinson, Kalirai, Simon, Kirby,
  VandenBerg, Mu{\~n}oz, Avila, Guhathakurta, \&
  Ferguson}]{Geha-2013-ApJ-771-29}
Geha, M., Brown, T.~M., Tumlinson, J., {et~al.} 2013, \apj, 771, 29

\bibitem[{Gonneau {et~al.}(2016)Gonneau, Lan\c{c}on, Trager, Aringer,
  Lyubenova, Nowotny, Peletier, Prugniel, Chen, Dries, Choudhury,
  Falc\'on-Barroso, Koleva, Meneses-Goytia, S\'anchez-Bl\'azquez, \&
  Vazdekis}]{Gonneau-2016-A&A-589-A36}
Gonneau, A., Lan\c{c}on, A., Trager, S.~C., {et~al.} 2016, \aap, 589, A36

\bibitem[{Gonneau {et~al.}(2017)Gonneau, Lan\c{c}on, Trager, Aringer, Nowotny,
  Peletier, Prugniel, Chen, \& Lyubenova}]{Gonneau-2017-A&A-601-A141}
Gonneau, A., Lan\c{c}on, A., Trager, S.~C., {et~al.} 2017, \aap, 601, A141

\bibitem[{Griest(1991)}]{Griest-1991-ApJ-366-412}
Griest, K. 1991, \apj, 366, 412

\bibitem[{Hargreaves {et~al.}(1994)Hargreaves, Gilmore, Irwin, \&
  Carter}]{Hargreaves-1994-MNRAS-269-957}
Hargreaves, J.~C., Gilmore, G., Irwin, M.~J., \& Carter, D. 1994, \mnras, 269,
  957

\bibitem[{Hawking(1971)}]{Hawking-1971-MNRAS-152-75}
Hawking, S. 1971, \mnras, 152, 75

\bibitem[{Hinton {et~al.}(2016)Hinton, Davis, Lidman, Glazebrook, \&
  Lewis}]{Hinton-2016-A&C-15-61}
Hinton, S.~R., Davis, T.~M., Lidman, C., Glazebrook, K., \& Lewis, G.~F. 2016,
  Astron. Comput., 15, 61

\bibitem[{Husser(2012)}]{Husser-2012-3DSDSP-UG-0}
Husser, T.-O. 2012, 3D-Spectroscopy of Dense Stellar Populations (G\"ottingen,
  Germany: Universit\"atsverlag G\"ottingen)

\bibitem[{Husser {et~al.}(2016)Husser, Kamann, Dreizler, Wendt, Wulff, Bacon,
  Wisotzki, Brinchmann, Weilbacher, Roth, \&
  Monreal-Ibero}]{Husser-2016-A&A-588-A148}
Husser, T.-O., Kamann, S., Dreizler, S., {et~al.} 2016, \aap, 588, A148

\bibitem[{Kamann {et~al.}(2016)Kamann, Husser, Brinchmann, Emsellem,
  Weilbacher, Wisotzki, Wendt, Krajnovi\'c, Roth, Bacon, \&
  Dreizler}]{Kamann-2016-A&A-588-A149}
Kamann, S., Husser, T.-O., Brinchmann, J., {et~al.} 2016, \aap, 588, A149

\bibitem[{Kamann {et~al.}(2018)Kamann, Husser, Dreizler, Emsellem, Weilbacher,
  Martens, Bacon, den Brok, Giesers, Krajnovi\'c, Roth, Wendt, \&
  Wisotzki}]{Kamann-2018-MNRAS-473-5591}
Kamann, S., Husser, T.-O., Dreizler, S., {et~al.} 2018, \mnras, 473, 5591

\bibitem[{Kamann {et~al.}(2013)Kamann, Wisotzki, \&
  Roth}]{Kamann-2013-A&A-549-A71}
Kamann, S., Wisotzki, L., \& Roth, M.~M. 2013, \aap, 549, A71

\bibitem[{{Knuth}(2006)}]{Knuth-arXiv-0605}
{Knuth}, K.~H. 2006, arXiv e-prints [\eprint[arXiv]{physics/0605197}]

\bibitem[{Koposov {et~al.}(2015)Koposov, Belokurov, Torrealba, \&
  Evans}]{Koposov-2015-ApJ-805-130}
Koposov, S.~E., Belokurov, V., Torrealba, G., \& Evans, N.~W. 2015, \apj, 805,
  130

\bibitem[{Koushiappas \& Loeb(2017)}]{Koushiappas-2017-PhRvL-119-041102}
Koushiappas, S.~M. \& Loeb, A. 2017, \prl, 119, 041102

\bibitem[{Kroupa(2001)}]{Kroupa-2001-MNRAS-322-231}
Kroupa, P. 2001, \mnras, 322, 231

\bibitem[{Li {et~al.}(2017)Li, Simon, Drlica-Wagner, Bechtol, Wang,
  Garcia-Bellido, Frieman, Marshall, James, Strigari, Pace, Balbinot, Zhang,
  Abbott, Allam, Benoit-L\'evy, Bernstein, Bertin, Brooks, Burke, Rosell, Kind,
  Carretero, Cunha, D'Andrea, da~Costa, DePoy, Desai, Diehl, Eifler, Flaugher,
  Goldstein, Gruen, Gruendl, Gschwend, Gutierrez, Krause, Kuehn, Lin, Maia,
  March, Menanteau, Miquel, Plazas, Romer, Sanchez, Santiago, Schubnell,
  Sevilla-Noarbe, Smith, Sobreira, Suchyta, Tarle, Thomas, Tucker, Walker,
  Wechsler, Wester, \& Yanny}]{Li-2017-ApJ-838-8}
Li, T.~S., Simon, J.~D., Drlica-Wagner, A., {et~al.} 2017, \apj, 838, 8

\bibitem[{Marsh \& Niemeyer(2019)}]{Marsh-2019-PhRvL-123-051103}
Marsh, D. J.~E. \& Niemeyer, J.~C. 2019, \prl, 123, 051103

\bibitem[{Martin {et~al.}(2018)Martin, Collins, Longeard, \&
  Tollerud}]{Martin-2018-ApJL-859-L5}
Martin, N.~F., Collins, M. L.~M., Longeard, N., \& Tollerud, E. 2018, \apjl,
  859, L5

\bibitem[{McConnachie(2012)}]{McConnachie-2012-AJ-144-4}
McConnachie, A.~W. 2012, \aj, 144, 4

\bibitem[{Monroy-Rodr\'{\i}guez \&
  Allen(2014)}]{MonroyRodriguez-2014-ApJ-790-159}
Monroy-Rodr\'{\i}guez, M.~A. \& Allen, C. 2014, \apj, 790, 159

\bibitem[{Navarro {et~al.}(1996)Navarro, Frenk, \&
  White}]{Navarro-1996-ApJ-462-563}
Navarro, J.~F., Frenk, C.~S., \& White, S. D.~M. 1996, \apj, 462, 563

\bibitem[{Navarro {et~al.}(1997)Navarro, Frenk, \&
  White}]{Navarro-1997-ApJ-490-493}
Navarro, J.~F., Frenk, C.~S., \& White, S. D.~M. 1997, \apj, 490, 493

\bibitem[{Ness {et~al.}(2015)Ness, Hogg, Rix, Ho, \&
  Zasowski}]{Ness-2015-ApJ-808-16}
Ness, M., Hogg, D.~W., Rix, H.-W., Ho, A. Y.~Q., \& Zasowski, G. 2015, \apj,
  808, 16

\bibitem[{O{\~n}orbe {et~al.}(2015)O{\~n}orbe, Boylan-Kolchin, Bullock,
  Hopkins, Kere{\v{s}}, Faucher-Gigu{\`e}re, Quataert, \&
  Murray}]{Onnorbe-2015-MNRAS-454-2092}
O{\~n}orbe, J., Boylan-Kolchin, M., Bullock, J.~S., {et~al.} 2015, \mnras, 454,
  2092

\bibitem[{Paxton {et~al.}(2011)Paxton, Bildsten, Dotter, Herwig, Lesaffre, \&
  Timmes}]{Paxton-2011-ApJS-192-3}
Paxton, B., Bildsten, L., Dotter, A., {et~al.} 2011, \apjs, 192, 3

\bibitem[{Paxton {et~al.}(2013)Paxton, Cantiello, Arras, Bildsten, Brown,
  Dotter, Mankovich, Montgomery, Stello, Timmes, \&
  Townsend}]{Paxton-2013-ApJS-208-4}
Paxton, B., Cantiello, M., Arras, P., {et~al.} 2013, \apjs, 208, 4

\bibitem[{Paxton {et~al.}(2015)Paxton, Marchant, Schwab, Bauer, Bildsten,
  Cantiello, Dessart, Farmer, Hu, Langer, Townsend, Townsley, \&
  Timmes}]{Paxton-2015-ApJS-220-15}
Paxton, B., Marchant, P., Schwab, J., {et~al.} 2015, \apjs, 220, 15

\bibitem[{Pe{\~n}arrubia {et~al.}(2012)Pe{\~n}arrubia, Pontzen, Walker, \&
  Koposov}]{Pennarrubia-2012-ApJ-759-L42}
Pe{\~n}arrubia, J., Pontzen, A., Walker, M.~G., \& Koposov, S.~E. 2012, \apjl,
  759, L42

\bibitem[{Renzini \& Ciotti(1993)}]{Renzini-1993-ApJ-416-L49}
Renzini, A. \& Ciotti, L. 1993, \apj, 416, L49

\bibitem[{Robin {et~al.}(2003)Robin, Reyl\'e, Derri\`ere, \&
  Picaud}]{Robin-2003-A&A-409-523}
Robin, A.~C., Reyl\'e, C., Derri\`ere, S., \& Picaud, S. 2003, \aap, 409, 523

\bibitem[{Robin {et~al.}(2004)Robin, Reyl\'e, Derri\`ere, \&
  Picaud}]{Robin-2004-A&A-416-157}
Robin, A.~C., Reyl\'e, C., Derri\`ere, S., \& Picaud, S. 2004, \aap, 416, 157

\bibitem[{Robitaille {et~al.}(2013)Robitaille, Tollerud, Greenfield,
  Droettboom, Bray, Aldcroft, Davis, Ginsburg, Price-Whelan, Kerzendorf,
  Conley, Crighton, Barbary, Muna, Ferguson, Grollier, Parikh, Nair, Guenther,
  Deil, Woillez, Conseil, Kramer, Turner, Singer, Fox, Weaver, Zabalza,
  Edwards, Bostroem, Burke, Casey, Crawford, Dencheva, Ely, Jenness, Labrie,
  Lim, Pierfederici, Pontzen, Ptak, Refsdal, Servillat, \&
  Streicher}]{Robitaille-2013-A&A-558-A33}
Robitaille, T.~P., Tollerud, E.~J., Greenfield, P., {et~al.} 2013, \aap, 558,
  A33

\bibitem[{Roth {et~al.}(2018)Roth, Sandin, Kamann, Husser, Weilbacher,
  Monreal-Ibero, Bacon, den Brok, Dreizler, Kelz, Marino, \&
  Steinmetz}]{Roth-2018-A&A-618-A3}
Roth, M.~M., Sandin, C., Kamann, S., {et~al.} 2018, \aap, 618, A3

\bibitem[{Salpeter(1955)}]{Salpeter-1955-ApJ-121-161}
Salpeter, E.~E. 1955, \apj, 121, 161

\bibitem[{Schlafly \& Finkbeiner(2011)}]{Schlafly-2011-ApJ-737-103}
Schlafly, E.~F. \& Finkbeiner, D.~P. 2011, \apj, 737, 103

\bibitem[{Schlegel {et~al.}(1998)Schlegel, Finkbeiner, \&
  Davis}]{Schlegel-1998-ApJ-500-525}
Schlegel, D.~J., Finkbeiner, D.~P., \& Davis, M. 1998, \apj, 500, 525

\bibitem[{Simon(2019)}]{Simon-2019-ARA&A-57-375}
Simon, J.~D. 2019, \araa, 57, 375

\bibitem[{Soto {et~al.}(2016)Soto, Lilly, Bacon, Richard, \&
  Conseil}]{Soto-2016-MNRAS-458-3210}
Soto, K.~T., Lilly, S.~J., Bacon, R., Richard, J., \& Conseil, S. 2016, \mnras,
  458, 3210

\bibitem[{{Stegmann} {et~al.}(2019){Stegmann}, {Capelo}, {Bortolas}, \&
  {Mayer}}]{Stegmann-arXiv-1910}
{Stegmann}, J., {Capelo}, P.~R., {Bortolas}, E., \& {Mayer}, L. 2019, arXiv
  e-prints [\eprint[arXiv]{1910.04793}]

\bibitem[{Stetson(1987)}]{Stetson-1987-PASP-99-191}
Stetson, P.~B. 1987, \pasp, 99, 191

\bibitem[{Weilbacher {et~al.}(2012)Weilbacher, Streicher, Urrutia, Jarno,
  P\'econtal-Rousset, Bacon, \& B\"ohm}]{Weilbacher-2012-SPIE-8451-84510B}
Weilbacher, P.~M., Streicher, O., Urrutia, T., {et~al.} 2012, \procspie, 8451,
  84510B

\bibitem[{Wheeler {et~al.}(2019)Wheeler, Hopkins, Pace, Garrison-Kimmel,
  Boylan-Kolchin, Wetzel, Bullock, Kere\v{s}, Faucher-Gigu\`ere, \&
  Quataert}]{Wheeler-2019-MNRAS-490-4447}
Wheeler, C., Hopkins, P.~F., Pace, A.~B., {et~al.} 2019, \mnras, 490, 4447

\bibitem[{Willman {et~al.}(2005)Willman, Blanton, West, Dalcanton, Hogg,
  Schneider, Wherry, Yanny, \& Brinkmann}]{Willman-2005-AJ-129-2692}
Willman, B., Blanton, M.~R., West, A.~A., {et~al.} 2005, \aj, 129, 2692

\bibitem[{Willman {et~al.}(2011)Willman, Geha, Strader, Strigari, Simon, Kirby,
  Ho, \& Warres}]{Willman-2011-AJ-142-128}
Willman, B., Geha, M., Strader, J., {et~al.} 2011, \aj, 142, 128

\bibitem[{Willman \& Strader(2012)}]{Willman-2012-AJ-144-76}
Willman, B. \& Strader, J. 2012, \aj, 144, 76

\bibitem[{Wolf {et~al.}(2010)Wolf, Martinez, Bullock, Kaplinghat, Geha,
  Mu{\~n}oz, Simon, \& Avedo}]{Wolf-2010-MNRAS-406-1220}
Wolf, J., Martinez, G.~D., Bullock, J.~S., {et~al.} 2010, \mnras, 406, 1220

\bibitem[{York {et~al.}(2000)York, Adelman, Anderson, Anderson, Annis, Bahcall,
  Bakken, Barkhouser, Bastian, Berman, Boroski, Bracker, Briegel, Briggs,
  Brinkmann, Brunner, Burles, Carey, Carr, Castander, Chen, Colestock,
  Connolly, Crocker, Csabai, Czarapata, Davis, Doi, Dombeck, Eisenstein,
  Ellman, Elms, Evans, Fan, Federwitz, Fiscelli, Friedman, Frieman, Fukugita,
  Gillespie, Gunn, Gurbani, de~Haas, Haldeman, Harris, Hayes, Heckman,
  Hennessy, Hindsley, Holm, Holmgren, Huang, Hull, Husby, Ichikawa, Ichikawa,
  Ivezic, Kent, Kim, Kinney, Klaene, Kleinman, Kleinman, Knapp, Korienek, Kron,
  Kunszt, Lamb, Lee, Leger, Limmongkol, Lindenmeyer, Long, Loomis, Loveday,
  Lucinio, Lupton, MacKinnon, Mannery, Mantsch, Margon, McGehee, McKay,
  Meiksin, Merelli, Monet, Munn, Narayanan, Nash, Neilsen, Neswold, Newberg,
  Nichol, Nicinski, Nonino, Okada, Okamura, Ostriker, Owen, Pauls, Peoples,
  Peterson, Petravick, Pier, Pope, Pordes, Prosapio, Rechenmacher, Quinn,
  Richards, Richmond, Rivetta, Rockosi, Ruthmansdorfer, Sandford, Schlegel,
  Schneider, Sekiguchi, Sergey, Shimasaku, Siegmund, Smee, Smith, Snedden,
  Stone, Stoughton, Strauss, Stubbs, SubbaRao, Szalay, Szapudi, Szokoly,
  Thakar, Tremonti, Tucker, Uomoto, Berk, Vogeley, Waddell, Wang, Watanabe,
  Weinberg, Yanny, \& Yasuda}]{York-2000-AJ-120-1579}
York, D.~G., Adelman, J., Anderson, J.~E., {et~al.} 2000, \aj, 120, 1579

\bibitem[{Zhu {et~al.}(2018)Zhu, Vasiliev, Li, \& Jing}]{Zhu-2018-MNRAS-476-2}
Zhu, Q., Vasiliev, E., Li, Y., \& Jing, Y. 2018, \mnras, 476, 2

\end{thebibliography}

    \begin{appendix}
    \section{Derivation of the limiting fraction of massive compact halo objects}
    \label{app:fMlim}
        \Citet{Binney-2008-GD-PU-2} give the following equations for the diffusion coefficients in the Fokker--Planck approximation of a subject star with mass~$m$ being perturbed by field stars of mass~$m_\mathrm{a}$:
        \begin{align}
            \begin{split}
            \mathrm{D}[\Delta E]
            &= m\Big(v\mathrm{D}[\Delta v_\parallel]\\
            &\qquad\;+ \frac{1}{2} \mathrm{D}[(\Delta v_\parallel)^2] + \frac{1}{2} \mathrm{D}[(\Delta\vec{v}_\perp)^2]\Big),
            \end{split}
            \\
            \mathrm{D}[\Delta v_\parallel] &= -\frac{4\pi G^2\rho(m+m_\mathrm{a}) \ln \Lambda}{\sigma^2} \mathcal{G}(X),\\
            \mathrm{D}[(\Delta v_\parallel)^2] &= \frac{4\sqrt{2}\pi G^2\rho m_\mathrm{a} \ln \Lambda}{\sigma} \frac{\mathcal{G}(X)}{X},\\
            \mathrm{D}[(\Delta\vec{v}_\perp)^2] &= \frac{4\sqrt{2}\pi G^2\rho m_\mathrm{a} \ln \Lambda}{\sigma} \frac{\mathrm{erf}\,X - \mathcal{G}(X)}{X},
        \end{align}
        where $v$ is the velocity of the subject stars, $\rho$ the background mass density of field stars, and $\sigma$ the velocity dispersion of field stars.
        The Coulomb logarithm~$\ln\Lambda$ is estimated with
        \begin{equation}
            \Lambda = \frac{b_\mathrm{max}v_\mathrm{typ}^2}{G(m+m_\mathrm{a})},
        \end{equation}
        where $b_\mathrm{max}$ is the maximum impact parameter and $v_\mathrm{typ}$ the typical relative velocity of the stars.
        $X$ and $\mathcal{G}(X)$ are defined as
        \begin{align}
            X &\coloneqq \frac{v}{\sqrt{2}\sigma},\\
            \mathcal{G}(X) &\coloneqq \frac{1}{2X^2} \Bigg(\mathrm{erf}\,X - \frac{2X}{\sqrt{\pi}} \exp(-X^2)\Bigg).
        \end{align}

        In our case, the `subject stars' are the stars and stellar remnants in the Eri~2 cluster, and the role of `field stars' is fulfilled by dark matter.
        We assume the dark matter consists of a mass fraction $f_\mathrm{M}$ of MACHOs with mass~$m_\mathrm{M}$ and cold, collisionless particles for the remaining fraction.
        The MACHOs heat up the star cluster, causing it to dissipate, while the low-mass particles have a cooling effect.
        \citet{Brandt-2016-ApJL-824-L31} argues this cooling effect is negligible compared to the heating effect from the MACHOs and discards it, but we keep it since it is not necessary to neglect it.

        We approximate the change in the star cluster's kinetic energy~$E_\mathrm{cl}$, which is the sum of the kinetic energies~$E$ of each star, with the relevant diffusion coefficient, which now consists of two terms:
        \begin{equation}
            \dot{E}_\mathrm{cl} \approx f_\mathrm{M}\mathrm{D}[\Delta E_\mathrm{cl}]\Big|_\mathrm{M} + (1-f_\mathrm{M})\mathrm{D}[\Delta E_\mathrm{cl}]\Big|_\mathrm{C}.
        \end{equation}
        The $|_\mathrm{M}$ and $|_\mathrm{C}$ indicate the diffusion coefficient in question has to be evaluated for the case of MACHOs and cold, collisionless particles, respectively.
        The linear pre-factors arise from the fact that the diffusion coefficient is linear in~$\rho$, which has to be modified to account for the mass fractions of the two kinds of dark matter particles.

        We can relate the kinetic energy to the gravitational potential energy~$U_\mathrm{cl}$ using the virial theorem $E_\mathrm{cl} = U_\mathrm{cl}/2$, and reorder to get
        \begin{equation}
            f_\mathrm{M} \approx \frac{\dot{U}_\mathrm{cl} - 2\mathrm{D}[\Delta E_\mathrm{cl}]\Big|_\mathrm{C}}{2\mathrm{D}[\Delta E_\mathrm{cl}]\Big|_\mathrm{M} - 2\mathrm{D}[\Delta E_\mathrm{cl}]\Big|_\mathrm{C}}.
        \end{equation}
        Like \citet{Brandt-2016-ApJL-824-L31}, we assume the star cluster is embedded in a dark-matter core, so the gravitational potential energy is given by
        \begin{equation}
            U_\mathrm{cl} = C - \frac{\alpha GM_\mathrm{cl}^2}{R_{\mathrm{h},\mathrm{cl}}} + \beta G\rho M_\mathrm{cl}R_{\mathrm{h},\mathrm{cl}}^2,
        \end{equation}
        where $M_\mathrm{cl}$ is the mass of the cluster and $\alpha$, $\beta$, and $C$ are constants with respect to the projected half-light radius~$R_{\mathrm{h},\mathrm{cl}}$.
        Assuming the dark-matter density and cluster mass are both constant, meaning the effect of the stars on the dark matter is negligible and no stars are kicked out of the cluster, the time derivative of the cluster potential is
        \begin{equation}
            \dot{U}_\mathrm{cl} = \Bigg(\frac{\alpha GM_\mathrm{cl}^2}{R_{\mathrm{h},\mathrm{cl}}^2} + 2\beta G\rho M_\mathrm{cl}R_{\mathrm{h},\mathrm{cl}}\Bigg)\dot{R}_{\mathrm{h},\mathrm{cl}}.
        \end{equation}

        In order to constrain the MACHO properties, we demand that the timescale for the dissipation of the cluster may not be shorter than the age of the cluster, otherwise we should expect to observe a much more diluted cluster.
        A natural choice for the dissipation timescale, considering the radius of the cluster is the only indication of dissipation that we can directly observe, is $R_{\mathrm{h},\mathrm{cl}}/\dot{R}_{\mathrm{h},\mathrm{cl}}$, so we define a limiting MACHO fraction
        \begin{equation}
            f_{\mathrm{M},\mathrm{lim}} \coloneqq \frac{\dot{U}_{\mathrm{cl},\mathrm{lim}} - 2\mathrm{D}[\Delta E_\mathrm{cl}]\Big|_\mathrm{C}}{2\mathrm{D}[\Delta E_\mathrm{cl}]\Big|_\mathrm{M} - 2\mathrm{D}[\Delta E_\mathrm{cl}]\Big|_\mathrm{C}},
        \end{equation}
        where
        \begin{equation}
            \dot{U}_{\mathrm{cl},\mathrm{lim}} \coloneqq \frac{\alpha GM_\mathrm{cl}^2 + 2\beta G\rho M_\mathrm{cl}R_{\mathrm{h},\mathrm{cl}}^3}{R_{\mathrm{h},\mathrm{cl}}t_\mathrm{cl}}.
        \end{equation}
        Since $f_\mathrm{M}$ increases with decreasing dissipation timescale, $f_{\mathrm{M},\mathrm{lim}}$ is an upper limit and we can exclude all combinations of parameters that give $f_\mathrm{M} > f_{\mathrm{M},\mathrm{lim}}$.
        The diffusion coefficients work out to be
        \begin{align}
            \begin{split}
                \mathrm{D}[\Delta E_\mathrm{cl}]\Big|_\mathrm{M}
                &= \frac{4\pi G^2\rho M_\mathrm{cl} \ln \Lambda_\mathrm{M}}{\sigma_\mathrm{DM}}\\
                &\quad\times \Bigg(\frac{m_\mathrm{M}\mathrm{erf}\,X}{\sqrt{2}X} - \frac{(m_*+m_\mathrm{M})\sigma_*\mathcal{G}(X)}{\sigma_\mathrm{DM}}\Bigg),
            \end{split}
            \\
            \mathrm{D}[\Delta E_\mathrm{cl}]\Big|_\mathrm{C} &\simeq \frac{4\pi G^2\rho M_\mathrm{cl} \ln \Lambda_\mathrm{C}}{\sigma_\mathrm{DM}} \Bigg(- \frac{m_*\sigma_*\mathcal{G}(X)}{\sigma_\mathrm{DM}}\Bigg),
        \end{align}
        where we have neglected the terms proportional to the mass of the cold, collisionless dark matter.
        $m_*$ is the mass of a typical star and $\sigma_*$ and $\sigma_\mathrm{DM}$ are the (three-dimensional) velocity dispersion of stars and dark matter particles, respectively.
        Following \citet{Brandt-2016-ApJL-824-L31}, we calculate the Coulomb logarithms using
        \begin{align}
            \Lambda_\mathrm{M} &= \frac{R_{\mathrm{h},\mathrm{cl}}\sigma_\mathrm{DM}^2}{G(m_*+m_\mathrm{M})},\\
            \Lambda_\mathrm{C} &= \frac{R_{\mathrm{h},\mathrm{cl}}\sigma_\mathrm{DM}^2}{Gm_*}
        \end{align}
        and we set
        \begin{equation}
            X = \frac{\sigma_*}{\sqrt{2}\sigma_\mathrm{DM}}.
        \end{equation}
        With the measurements and assumptions of this paper, we have $\ln \Lambda_\mathrm{C} \approx 1$ and $\ln \Lambda_\mathrm{M}$ can reach up to ${\sim}12$ depending on $m_\mathrm{M}$.  For $m_\mathrm{M} = 100\,M_\mathrm{\sun}$, it is ${\sim}8$.
    \end{appendix}
\end{document}